\documentclass[onecolumn,10pt]{IEEEtran}
\usepackage{amsmath}
\usepackage{amsthm}
\usepackage{amssymb}
\usepackage{graphicx}
\usepackage{caption}
\usepackage{algorithm}
\usepackage{algpseudocode}
\usepackage{cite}
\usepackage{xcolor}
\title{}
\author{}
\date{}

\newtheorem{definition}{Definition}
\newtheorem{theorem}{Theorem}
\newtheorem{construction}{Construction}
\newtheorem{example}{Example}
\newtheorem{lemma}{Lemma}
\newtheorem{remark}{Remark}
\newtheorem{corollary}{Corollary}
\newcommand{\tabincell}[2]{\begin{tabular}{@{}#1@{}}#2\end{tabular}}
\begin{document}
	\title{A New Construction Structure on Coded Caching with Linear Subpacketization: Non-Half-Sum Latin Rectangle }
	\author{Yongcheng Yang, Minquan Cheng,~\IEEEmembership{Member,~IEEE,} 
		Kai Wan,~\IEEEmembership{Member,~IEEE,}  and Giuseppe Caire,~\IEEEmembership{Fellow,~IEEE}
		\thanks{Y. Yang and M. Cheng are with Guangxi Key Lab of Multi-source Information Mining $\&$ Security, Guangxi Normal University,
			Guilin 541004, China (e-mail: yongchengyang1@163.com, chengqinshi@hotmail.com).}
		\thanks{K.~Wan is with the School of Electronic Information and Communications,
			Huazhong University of Science and Technology, 430074  Wuhan, China,  (e-mail: kai\_wan@hust.edu.cn).}
		\thanks{G. Caire is with the Electrical Engineering and Computer Science Department, Technische Universit\"{a}t Berlin,10587 Berlin, Germany (e-mail: caire@tu-berlin.de).}
	}

	\maketitle
	\begin{abstract}
		Coded caching is recognized as an effective method for alleviating network congestion during peak periods by leveraging local caching and coded multicasting gains. The key challenge in designing  coded caching schemes lies in simultaneously achieving low subpacketization and low transmission load. Most existing schemes require exponential or polynomial subpacketization levels, while some linear subpacketization schemes often result in excessive transmission load. Recently, Cheng et al. proposed a construction framework for linear coded caching schemes called Non-Half-Sum Disjoint Packing (NHSDP), where the subpacketization equals the number of users $K$. This paper introduces a novel combinatorial structure, termed the Non-Half-Sum Latin Rectangle (NHSLR), which extends the framework of linear coded caching schemes from $F=K$ (i.e., the construction via NHSDP) to a broader scenario with $F=\mathcal{O}(K)$. By constructing NHSLR, we have obtained a new class of coded caching schemes that achieves linearly scalable subpacketization, while further reducing the transmission load compared with the NHSDP scheme. Theoretical and numerical  analyses demonstrate that the proposed schemes not only achieves lower transmission load than existing linear subpacketization schemes but also approaches the performance of certain exponential subpacketization schemes.	
	\end{abstract}
	
	\begin{IEEEkeywords}
		Coded caching, placement delivery array,  linear subpacketization, transmission load, Non-Half-Sum Latin Rectangle
	\end{IEEEkeywords}
	
	\section{Introduction}
	Recently, the proliferation of wireless devices has dramatically increased network traffic, driven by activities such as multimedia streaming, web browsing, and social networking. This surge in traffic, combined with its high temporal variability, often results in network congestion during peak hours and inefficient resource utilization during off-peak periods. Caching emerges as an effective solution to mitigate peak-time load by leveraging distributed memory across the network to store content during low-traffic periods. Traditional uncoded caching strategies typically rely on predicting user demands to formulate prefetching plans, thereby achieving a “local caching gain” that grows with the size of local storage \cite{BGW}. Maddah-Ali and Niesen (MN)  in \cite{MN} introduced the concept of coded caching, showing that it not only provides local caching gain but also enables a global caching gain. This global gain arises from the ability of coded transmissions to simultaneously satisfy multiple users, and it scales with the total memory available throughout the network.
	
	A $(K,M,N)$ coded caching system consists of a central server storing $N$ files of equal size and $K$ users, each equipped with a cache capable of storing up to $M$ files. The server communicates with the users over a  error-free shared-link. An 
	$F$-division $(K,M,N)$ coded caching scheme operates in two phases: the placement phase during off-peak hours and the delivery phase during peak hours. In the placement phase, the server splits each file into $F$ equal-sized packets and distributes certain packets to each user's cache without knowledge of future user requests. If the packets are stored directly in users' caches without any coding, the process is termed uncoded placement; otherwise, it is called coded placement. The value $F$ is known as the subpacketization. During the delivery phase, each user randomly requests one file. Based on the users’ requests and their cached content, the server transmits carefully designed coded packets over the shared link, allowing every user to reconstruct their requested file. The transmission load $R$ refers to the normalized amount of data transmitted in the worst-case scenario across all possible demands.                                        
	
	Maddah-Ali and Niesen (MN) were the first to propose a coded caching scheme (termed the MN scheme)  \cite{MN}. This scheme has optimal transmission load under uncoded placement when $K\le N$ \cite{WTP2016,WTP2020}, and is order optimal within a factor of $4$ \cite{GR}. In the event of $K > N$, Yu et al. refined the MN scheme by eliminating redundant transmissions, thereby obtaining an optimal scheme under uncoded placement \cite{YMA2018}. This was shown to be order optimal within a factor of $2$ in \cite{YMA2019}. Jin et al. in \cite{JCLC} derived the average load under conditions such as non-uniform file popularity and various demands. They demonstrated that when $K$ is less than or equal to $N$ with uniform popularity, the minimum load is precisely that of the MN scheme. However, the subpacketization $F={K\choose KM/N}$ of the MN scheme grows exponentially with the number of users $K$, which makes it extremely difficult in practical engineering realizations. Therefore, designing coded caching schemes with low subpacketization has become a meaningful and active area of research. Among various approaches, the grouping method introduced in  \cite{SJTLD,CJWY} is widely considered as the most effective strategies for reducing subpacketization. Nonetheless, this method often results in a rapidly increasing transmission load.
	
	More recently, the authors in \cite{YCTC} introduced a novel combinatorial characterization known as the Placement Delivery Array (PDA) to study coded caching schemes with low subpacketization. They also showed that the MN scheme can be represented by a particular type of PDA, known as an MN PDA. Numerous schemes with lower subpacketization than the MN scheme are obtained by constructing PDAs in  \cite{CJWY, YCTC,CJYT,CJTY,CWZW,WCWG,WCWL,WCCLS,ZCW,LC,PKB,WCLC,AST}. There are some other characterizations including linear block codes \cite{TR}, $(6,3)$-free hypergraphs \cite{SZG}, $(r,t)$ Ruzsa-Szem\'{e}redi graphs \cite{STD}, strong edge colorings of bipartite graphs \cite{YTCC}, cross resolvable designs \cite{KMR}, projective geometries \cite{CKSM}, and combinatorial designs \cite{ASK}, as listed in Table~\ref{tab-Schemes}. As illustrated in Table \ref{tab-Schemes}, the schemes from \cite{STD} and the first scheme in \cite{XXGL} have been omitted due to the absence of explicit construction.  
	\begin{table}[http!]
		\renewcommand{\arraystretch}{2}
		\setlength\tabcolsep{1pt} 
		\centering
		\caption{The existing coded caching schemes where $K,k,t,m,H,a,z,r \in \mathbb{Z}^{+}$, $\left[k \atop t \right]_q=\frac{(q^{k}-1)\dots(q^{k-t+1}-1)}{(q^{t}-1)\dots(q-1)}$, $\left \langle K \right \rangle_{t}=K \mod t$.}
		\label{tab-Schemes}
		\begin{tabular}{|c|c|c|c|c|c|}\hline
			Schemes & Number of Users & Memory ratio  & Load    & Subpacketization &Constraint\\ \hline
			
			MN Scheme  \cite{MN}&$K$               & $\frac{t}{K}$& $\frac{K-t}{t+1}$& ${K\choose t}$         &   \\ \hline
			\tabincell{c}{WCLC  \cite{WCLC}}&$\binom{m}{z}k^z$               & $1-\left(\frac{k-t}{k}\right)^{z}$& $(\frac{k-t}{\lfloor\frac{k-1}{k-t}\rfloor})^z$ & ${\lfloor\frac{k-1}{k-t}\rfloor}^zk^{m-1}$         &\tabincell{c}{$1\leq t < k$,\\$1\leq z \leq m$}    \\ \hline
			
			
			YTCC  \cite{YTCC}&$\binom{H}{a}$& $1-\frac{\binom{a}{r}\binom{H-a}{b-r}}{\binom{H}{b}}$
			&\tabincell{c}{$\frac{\binom{H}{a+b-2r}}{\binom{H}{b}}\cdot$ \\ $\min\{\binom{H-a-b+2r}{a-r},$\\ $\binom{a+b-2r}{a-r}\}$}
			& $\binom{H}{b}$&\tabincell{c}{$r< a < H,$\\ $r<b<H,$\\$a+b \leq H+r$}          \\ \hline
			
			WCCLS  \cite{WCCLS}  &$q^m$ &$1-\frac{\binom{m}{w}(q-1)^w}{q^m}$ &$\frac{\binom{m}{w}(q-1)^w}{q^{m-w}}$&$q^m$&$m,w \in \mathbb{Z}^{+},m<w$
			\\ \hline
			
			\tabincell{c}{CKSM  1 \cite{CKSM}}
			&\tabincell{c}{$\frac{1}{t!}q^{\frac{t(t-1)}{2}}\cdot$\\ $\prod \limits_{i=0}^{t-1}\left[k-i \atop 1 \right]_q$}
			&\tabincell{c}{$1-q^{mt}\cdot$ \\  $\prod \limits_{i=0}^{m-1}\frac{\left[k-t-i \atop 1\right]_q}{\left[k-i \atop 1\right]_q}$}& \tabincell{c}{$\frac{m!q^{mt}}{(m+t)!}q^{\frac{t(t-1)}{2}}\cdot$\\  $\prod \limits_{i=0}^{t-1}\left[k-m-i \atop 1\right]_q$}&
			\tabincell{c}{$\frac{1}{m!}q^{\frac{m(m-1)}{2}}\cdot$\\ $\prod \limits_{i=0}^{m-1}\left[k-i \atop 1 \right]_q$}&
			\tabincell{c}{$m+t \leq k$,\\  prime power}\\ \cline{1-5}
			
			\tabincell{c}{CKSM  2 \cite{CKSM}}&
			$\left[ k \atop t\right]_q$&$1-\frac{\left[k-t \atop m \right]_q}{\left[k \atop m+t\right]_q}$                  &$\frac{\left[k \atop m\right]_q}{\left[k \atop m+t\right]_q}$      &$\left[k \atop m+t\right]_q$&$2\leq q$ \\ \hline
			
			\tabincell{c}{ASK  1 \cite{ASK}}&
			$q^2+q+1 $&$\frac{q^2}{q^2+q+1}$&$1$&$q^2+q+1$& prime power     \\  \cline{1-5}
			\tabincell{c}{ASK  2  \cite{ASK} }
			&$q^2$       & $\frac{q-1}{q}$              &$\frac{q}{q+1}$      &$q^2+q$   & $2\leq q$         \\ \hline

			\tabincell{c}{ZCW  \cite{ZCW} } &$2^{m}$ &$1-\frac{\binom{m}{\omega}}{\sum_{i=0}^{\omega}\binom{m}{i}}$  &$\frac{\binom{m}{\omega}2^{m-\omega}}{\sum_{i=0}^{\omega}\binom{m}{i}}$ & $\sum_{i=0}^{\omega}\binom{m}{i}$  &  $\omega <  m$ \\ \hline

			& & &$\frac{(K-t)(K-t+1)}{2K}$ &$K$ &\tabincell{c}{ $(K-t+1)|K$ \\ or $K-t=1$} \\ \cline{4-6}
			WCWL  \cite{WCWL} & \tabincell{c}{$K$}  &$\frac{t}{K}$  &$\frac{K-t}{2\lfloor \frac{K}{K-t+1}\rfloor+1}$ &$\left(2\lfloor \frac{K}{K-t+1}\rfloor+1\right)K$ &\tabincell{c}{$\left \langle K \right \rangle_{K-t+1}=K-t$ } \\ \cline{4-6}
			&   & &$\frac{K-t}{2\left\lfloor \frac{K}{K-t+1}\right\rfloor}$ &$2\left\lfloor \frac{K}{K-t+1}\right\rfloor K$ &\tabincell{c}{ $\mbox{otherwise}$ } \\ \hline
			XXGL  \cite{XXGL} &$K$ &$\frac{K-2}{K}$ &$\frac{K-1}{K}$ &$K$ & $K \in \mathbb{N}^{+}$ \\ \hline
			AST  \cite{AST} &$2^{r}k$  &$1-\frac{r+1}{2^{r}}+\frac{r}{2^{r}k}$ &$\frac{k(r+1)-r}{2^{r}}$  &$2^{r}k=K$   &$r,k \in \mathbb{N}^{+}$ \\ \hline
			MR  \cite{MR}  &$K$  &$\frac{t}{K}$  &\small$\left\lceil\frac{K(K-t)}{2+\left \lfloor\frac{t}{K-t+1}\right \rfloor+\left \lfloor\frac{t-1}{K-t+1}\right\rfloor}\right\rceil \cdot \frac{1}{K}$ &$K$  & \\ \hline
			CWWC  \cite{CWWC}  &$q^n$  &$1-\frac{2^n\lfloor\frac{q-1}{2}\rfloor^n}{q^n}$  &\small${\left\lfloor \frac{q-1}{2} \right\rfloor}^n$ &$q^n$  & $q^n$ is odd \\ \hline
		\end{tabular}
	\end{table}
	
	As outlined in Table~\ref{tab-Schemes}, the schemes proposed in \cite{CJWY,MN,YCTC,CJYT,TR,SZG,WCWG,WCLC,CWZW} exhibit flexibility in terms of user numbers, support large memory regimes, and achieve low transmission loads. However, they suffer from high subpacketization levels, which grow exponentially or sub-exponentially with the number of users. In contrast, the schemes in \cite{YTCC,ZCW,WCWL,SS,AST,WCCLS,CKSM,ASK,XXGL,MR} are designed to achieve low subpacketization, with growth rates that are polynomial or even linear in the number of users. Nevertheless, many of these low subpacketization schemes impose constraints on system parameters. For instance, the schemes in \cite{CKSM,YTCC,WCCLS} are only applicable to specific numbers of users, which are often defined as combinations, powers, or their products, and require particular memory ratios. The schemes in \cite{ZCW,ASK,XXGL} operate under large memory ratios, typically approaching $1$, while those in \cite{AST,CKSM} support extreme memory ratios close to either $0$ or $1$. Of particular interest are the schemes in \cite{MR,WCWL}, which achieve linear subpacketization with respect to the number of users, while still accommodating a flexible user count and large memory regimes. However their transmissions are too large. To establish a constructing framework for the coded caching schemes linear with $K=F$, Cheng et al. in \cite{CWWC} recently introduced a novel combinatorial structure termed NHSDP, which combine placement and delivery strategies into a single combinatorial condition. By constructing NHSDP, they obtained a class of linear subpacketization (i.e., $F=K$) coded caching schemes which attain a lower transmission load compared to existing schemes at the same subpacketization level.
	\subsection{Research Motivation and Contribution}
	In this paper, we focus on studying coded caching schemes with linear subpacketization. A new construction framework for coded caching schemes with $F=\mathcal{O}(K)$ is proposed by relaxing the limitation $F=K$ of the NHSDP. This framework, called the Non-Half Sum Latin Rectangle (NHSLR), can be used to generate schemes that achieve a larger coded caching gain and a wider memory range than the NHSDP scheme. The main contributions of this paper are as follows:
	
	
	\begin{itemize}
		\item We propose a new combinatorial structure, termed NHSLR, which unifies placement and transmission policies into a single condition when $F=\mathcal{O}(K)$. 
		A $g\times b$ Latin Rectangle $\mathbf{D}$ over $\mathbb{Z}_v$ where $v$ is odd positive is a $(v, g, b)$ NHSLR if any two different rows of $\mathbf{D}$ satisfy some conditions of the non-half-sum (the sum is divided by $2$ over $\mathbb{Z}_v$) property defined in Definition \ref{def-NHSLR}. 
		Give a $(v,g,b)$ NHSLR with positive integers $g$ and $b$, we can construct a $(K=v,M,N)$ coded caching scheme with memory ratio  $\frac{M}{N}=1-\frac{b}{v}$, subpacketization $F=gK$, and transmission load $R=\frac{b}{g}$. If the number of users is not odd, a virtual user can be introduced so that the effective number of users is $K'=K+1$, and then a scheme based on the $(K+1, g, b)$ NHSLR can be constructed. It is worth noting that NHSDP can be used to construct NHSLR, but not vice versa. This implies that combinatorial structure is less than that of the NHSDP. In addition, compared to the scheme generated by directly replicating the NHSDP scheme, the NHSLR scheme can more fully utilize cached information, thereby to realize larger multicast gains. 
		\item By carefully designing the diagonal matrix $\mathbf{X}$, we construct a class of NHSLRs. 
		The core of this construction lies in the fact that the diagonal entries of $\mathbf{X}$ can linearly represent the entries of $\mathbf{D}$ while also representing the elements of the half-sum of any two rows of $\mathbf{D}$.  Moreover, each entry in the half-sum of any two distinct rows of $\mathbf{D}$ does not occur in either of the two original rows. By constructing $\mathbf{X}$, the following results can be obtained.
		\begin{itemize}
			\item For any positive integer $n$ and arbitrary positive integers $m_1, m_2, \ldots, m_n$, we obtain an $(v,2^n,\prod_{i=1}^{n} m_i)$ NHSLR, which yields a $(K=v, M, N)$ coded caching scheme with memory ratio $M/N = 1 - \frac{\prod_{i=1}^{n} m_i}{v}$, subpacketization $F = vg$, coded caching gain $g = 2^n$, and transmission load $R = \frac{\prod_{i=1}^{n} m_i}{2^n}$, where $v$ is any odd integer satisfying $v \geq \prod_{i=1}^{n}{(m_i+1)}$.
			\item In particular, when the parameters are chosen uniformly as $m_1 = m_2 = \cdots = m_n = \lfloor q - 1 \rfloor$ with $q = v^{1/n}$ where $q$ is a real number, the resulting scheme has memory ratio $M/N = 1 - \frac{\lfloor q - 1 \rfloor^n}{q^n}$, subpacketization $F = vg$, coded caching gain $g = 2^n$, and transmission load $R = (\frac{\lfloor q-1 \rfloor}{2})^n$.
		\end{itemize}
		
		\item Theoretical analysis and numerical comparisons show that the proposed schemes achieve a lower transmission load than existing schemes with linear subpacketization. Compared to schemes with polynomial subpacketization, our schemes not only attain a lower subpacketization level but also achieve a lower transmission load in some cases. Moreover, the transmission load of our schemes is close to that of existing schemes with exponential subpacketization for some memory ratio and users. 
	\end{itemize}
	\subsection{Organizations and Notations}
	The rest of this paper is organized as follows. In Section~\ref{sec-2}, we introduce the coded cache systems, concepts of PDAs, and their relationships. In Section~\ref{sec-3}, we introduced the definition of Non-Half-Sum Latin Rectangle (NHSLR) and construction of PDA via NHSLR. In Section~\ref{sec-4}, we provide a class of constructions for NHSLR. Section~\ref{sec-5} analyzes the performance of the proposed scheme. In Section~\ref{sec-6}, we provide the proof of Theorem \ref{th-main-PDA}. Finally, we conclude the paper in Section~\ref{sec-7}.

	\textit{Notation:} In this paper, we will use the following notations. Let bold capital letter, bold lowercase letter, and curlicue letter  denote array, vector, and set respectively; let $|\mathcal{A}|$ denote the cardinality of the set $\mathcal{A}$; define  $[a]=\{1,2,\ldots,a\}$ and $[a:b]$ is the set $\{a,a+1,\dots,b-1,b\}$; $\lfloor q\rfloor $ denotes the largest integer not greater than $q$; define $\max{\{\mathcal{T}\}}$ to denote the largest entrie in the set $\mathcal{T}$. We define that $\left[k \atop t \right]_q=\frac{(q^{k}-1)\dots(q^{k-t+1}-1)}{(q^{t}-1)\dots(q-1)}$, $\left \langle K \right \rangle_{t}=K\mod t$; $\mathbb{Z}_v$ is the ring of integer residues modulo $v$.
	
	\section{Preliminaries}\label{sec-2}
	In this section, we will introduce a coded caching system, placement delivery array and their relationship respectively.
	\subsection{Coded Caching System}
	As illustrated in Figure~\ref{system-model}, a $(K,M,N)$ coded cache system consists of a server with $N$ equal-sized files $\mathcal{W}=\{W_n \mid n\in[N]\}$ and $K$ users storing at most $M$ file sizes where $0\leq M\leq N$. The servers are connected to the users through error-free shared-link.  An $F$-division $(K,M,N)$ coded caching scheme consists of the following two phases:
	\begin{figure}[h]
		\centering
		\includegraphics[height=5cm,width=7cm]{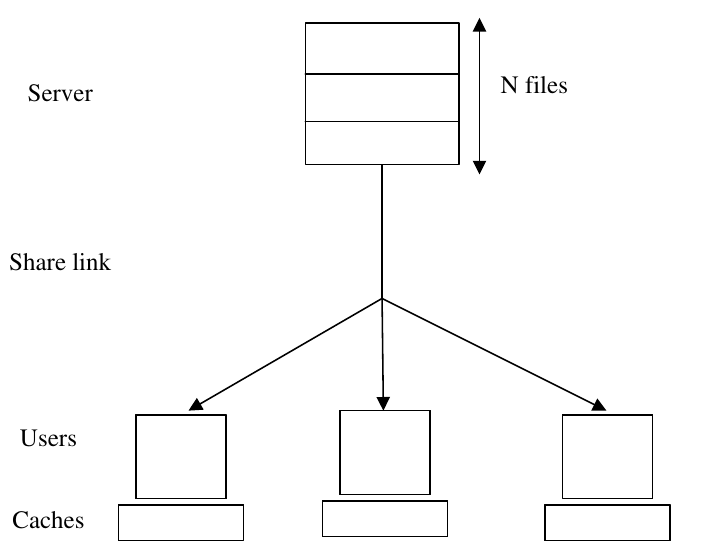}
		\caption{$(K,M,N)$ caching system}
		\label{system-model}
	\end{figure}
	\begin{itemize}
		\item {\bf Placement phase:} During off-peak traffic times, each file is divided into $F$ equal-sized packets, i.e., $W_n=\{W_{n,j}\mid j\in[F]\}$ for $n \in [N]$, and some packets are placed into each user's cache without knowledge of future demands. Let $\mathcal{Z}_{k}$ denote the packets cached by user $k$. This placement strategy is referred to as uncoded placement. In contrast, if coding is applied during placement, it is called coded placement.
		
		\item {\bf Delivery phase:} During the peak traffic times, each user randomly requests one file in  $\mathcal{W}$. Denote the requested vector by $\mathbf{d}=(d_{1},d_{2},\ldots,d_{K})$, i.e., user $k$ requests file $W_{d_{k}}$, where $k \in \mathcal{K},d_{k} \in [N]$. After receiving the request vector ${\bf d}$, the server broadcasts at most $R_{\bf d}F$ XOR of coded packets to the users such that each user is able to decode its requested file.
	\end{itemize}
	
	In this paper, we focus on the normalized amount of transmission for the worst-case over all possible demands which is defined as follows.
	\begin{equation}
		\label{eq:def of load}
		R=\max\{R_{\bf d} \mid {\bf d}\in[N]^K\}.
	\end{equation}
	The first well-known scheme, proposed in \cite{MN} and referred to as the MN scheme, achieves the minimum load under uncoded placement \cite{WTP2016,JCLC,YMA2018,WTP2020}. However, it suffers from high subpacketization, which grows exponentially with the number of users.
	
	\subsection{Placement Delivery Array}
	In order to study schemes with low subpacketization, the authors in \cite{YCTC} propose a novel combinatorial structure called the  placement delivery array, defined as follows.
	\begin{definition}[PDA,\cite{YCTC}]\rm
		\label{def-PDA}
		For positive integers $K$, $F$, $Z$ and $S$, an $F \times K $ array $\mathbf{P}=(p_{j,k})_{j \in [F],k \in [K]}$, composed of symbol $``*"$ and $[S]$, is called a $(K,F,Z,S)$ PDA if the following conditions hold:
		\begin{enumerate}
			\item [C$1$.] Each column has exactly $Z$ stars.
			\item [C$2$.] Each integer in $[S]$ occurs at least once.
			\item [C$3$.] For any two distinct entries $p_{j_{1},k_{1}}$ and $p_{j_{2},k_{2}}$, $p_{j_{1},k_{1}} = p_{j_{2},k_{2}}=s$ if only if
			\begin{enumerate}
				\item [a).] $j_{1} \neq j_{2}, k_{1} \neq k_{2}$, i.e., they lie in distinct rows and distinct columns; 
				\item [b).] $ p_{j_{1}, k_{2}}= p_{j_{2}, k_{1}}=*$, i.e., the corresponding $2 \times 2$ subarray formed by rows $j_{1}, j_{2}$ and columns  $k_{1}, k_{2}$ must be one of the following form,
				\begin{align}\label{eq-form-pda}
					\left(\begin{array}{cc}
						s &*\\
						* & s
					\end{array}\right)\ \ \ \ \textrm{or}\ \ \ \ 
					\left(\begin{array}{cc}
						* & s\\
						s &*
					\end{array}\right).
				\end{align}
			\end{enumerate}
		\end{enumerate}\hfill $\square$
	\end{definition}
	Let us take the following example to illustrate the concept of PDA. 
	When $(K,F,Z,S)=(4,4,2,4)$, let us consider the following array
	\begin{eqnarray}\label{ex-pda}
		\mathbf{P}=\left(
		\begin{array}{ccccc}
			*&*&1&4\\
			1&*&*&2\\
			3&2&*&*\\
			*&4&3&*\\
		\end{array}
		\right).
	\end{eqnarray}
	Clearly, each column contains exactly $Z = 2$ stars, and each integer appears at least once in $\mathbf{P}$. Therefore, conditions C$1$–C$2$ hold. When $s = 1$, we can check that $p_{1,3} = p_{2,1} = 1$ and $p_{1,1} = p_{2,3} = *$, which satisfies condition C$3$. Similarly, we can verify that all integer entries satisfy condition C$3$. Hence, $\mathbf{P}$ is a $(4,4,2,4)$ PDA.  
	
	Given a $(K,F,Z,S)$ PDA $\mathbf{P}$, we can obtain a coded caching scheme in the following way. The columns and rows of $\mathbf{P}$ represent the user  and file packet labels respectively. We set that user $k$ stores the $j$-th packet of every file only if $p_{j,k}=*$. By Condition C1, each user caches exactly $M=NZ/F$ files, i.e., the memory ratio of $M/N = Z/F$. Clearly,  $p_{j,k}\neq *$ means that user $k$ does not cache the $j$-th packet of each file. For each integer $s\in[S]$,  the server broadcasts a coded multicast message (i.e., the XOR of all requested packets corresponding to $s$) to all users at times slot $s$. By the condition C$3$, each severed user at time slot $s$ can decode its required packets since it caches all the other packets. By Condition C2, there are exactly $S$ time slots in the delivery phase. In addition, each time slot the coded packet sent by the sever has the same size of the packet. Assume that $s$ occurs $g$ times. Then the server severs exactly $g$ users, i.e., the coded caching gain is $g$. So the transmission load is $R=S/F$. Then the following result can be obtained.   
	
	\begin{lemma}[Scheme via PDA,\cite{YCTC}]\rm
		\label{le-PDA}
		Given a $(K,F,Z,S)$ PDA, there exists an $F$-division coded caching scheme for the $(K,M,N)$ coded caching system with memory ratio $M/N=Z/F$, subpacketization $F$, and load $R=S/F$.
		\hfill $\square$
	\end{lemma}
	
	
	There are many existing PDA schemes such as \cite{CJWY, CJYT, CJTY, CWZW, ZCJ, ZCW, YCTC, YTCC, SZG, MW}, but most of them require that the subpacketizations increase either exponentially or polynomially with the number of users. Recently, the authors in \cite{CWWC} proposed a linear subpacketization coded caching scheme construction framework called Non-half-sum disjoint packing (NHSDP). 
	

	\begin{definition}[Non-half-sum disjoint packing, NHSDP \cite{CWWC}]\label{def-NHSDP}\rm
		For any positive odd integer $v$, a pair $(\mathbb{Z}_v,\mathfrak{D})$ where $\mathfrak{D}$ consists of $z$ $g$-subsets of $\mathbb{Z}_v$ is called $(v,g,z)$ non-half-sum disjoint packing if the following conditions hold.
		\begin{itemize}
			\item  The intersection of any two different entries in $\mathfrak{D}$ is empty;  
			\item  For  each $\mathcal{D}\in \mathfrak{D}$, the half-sum of any two different entries in $\mathcal{D}$ (i.e., the sum of the two entries divided by $2$) does not appear in any block of $\mathfrak{D}$. \hfill $\square$
		\end{itemize}
	\end{definition}

	Using NHSDP, we can obtain the following result.
	\begin{lemma}[\cite{CWWC}]\rm
		\label{le-NHSDPPDA}
		Given a $(v,g,z)$ NHSDP, there exists a $(v,v,v-zg,zv)$ PDA which realizes a $(K=v,M,N)$ coded caching scheme with memory ratio $M/N=1-zg/v$, coded caching gain $g$, subpacketization $F=v$, and  transmission load $R=z$.    \hfill $\square$
	\end{lemma}
	
	By Lemma~\ref{le-NHSDPPDA}, NHSDP can be used to construct a linear subpacketization coded caching scheme with $F=K$. In the following, we relax the constraint $F=K$ in the NHSDP scheme and propose a new construction framework that achieves $F = \mathcal{O}(K)$. 
	\section{Construction of PDA via Non-half-sum Latin Rectangle}
	\label{sec-3}
	In this section, we first introduce the concept of Non-Half-Sum Latin Rectangle (NHSLR) to construct PDA, then explore the relationship between NHSDP and NHSLR. 
	
	According to Condition C$3$-a) in Definition~\ref{def-PDA}, the authors in \cite{CWWC} constructed a PDA with $F=K$ using a single Latin square (i.e., based on the Latin square construction, they derived a novel combinatorial construction framework for PDAs, called Non-Half-Sum Disjoint Packing). Naturally, it is interesting whether we can use $g$ Latin squares to study the PDA with the subpacketization $F=gK$ for some integer $g$, i.e., $F=\mathcal{O}(K)$. However, directly applying the NHSDP framework to each of the $g$ Latin squares would result in $g$ independent PDAs, where multicast occurs only within a single Latin square, failing to enhance multicast gain. To improve the multicast gain, the PDA structure should be considered as an integrated whole. So we changed the PDA structure in~\cite{CWWC} to generate multicast opportunities across different Latin square. This requires a new combinatorial structure to characterize the constraints of integer selection in the first row of each Latin square. Therefore, we further propose a new combinatorial structure called Non-Half-Sum Latin Rectangle (NHSLR), whose properties correspond precisely to the conditions required for constructing a PDA. Specifically, the Latin property of the NHSLR ensures the row and column uniqueness required by the PDA, while its non-half-sum property guarantees that the cross positions specified in Condition~C$3$ are filled with star. Now, let us introduce its formal definition first. 
	\begin{definition}[Non-half-sum latin rectangle, NHSLR]\label{def-NHSLR}\rm
		For any positive odd integer $v$ and positive integers $g$ and $b$, a $g \times b$ array  $\mathbf{D}$ over $\mathbb{Z}_v$ is called a $(v,g,b)$ NHSLR if the following conditions hold:
		\begin{itemize}
			\item Each integer in $\mathbb{Z}_v$ appears in each row and each column at most once (i.e., Latin property).
			\item For any two distinct row vectors of $\mathbf{D}$, say $\mathbf{d}$ and $\mathbf{d}'$, no integer of the half-sum vector
			$\frac{1}{2}(\mathbf{d}+ \mathbf{d}')$ 
			appears in $\mathbf{d}$ or $\mathbf{d}'$ (i.e., non-half-sum property).\hfill $\square$
		\end{itemize}
	\end{definition}
	Let us take the following example with $v = 7$ to further explain the concept of NHSLR.
	\begin{example}\rm
		\label{ex-NHSLR}
		When $(v,g,b)=(7,3,4)$, let us consider the following array
		\begin{align}\rm\label{eq-ex-1}
			\mathbf{D}=\begin{pmatrix}
				\mathbf{d}_1\\
				\mathbf{d}_2\\
				\mathbf{d}_3\\
			\end{pmatrix}=\begin{pmatrix}
				1&2&3&4\\
				2&1&4&6\\
				4&5&2&1
			\end{pmatrix}.
		\end{align}
		We can observe that each integer in $\mathbb{Z}_7$ appears at most once in each row and column of $\mathbf{D}$; hence, the first condition in Definition~\ref{def-NHSLR} is satisfied. All the corresponding half-sum vectors are given as follows
		\begin{align*}
			\mathbf{d}_{12}=\frac{\mathbf{d}_1+\mathbf{d}_2}{2}=(\frac{1+2}{2},\frac{2+1}{2},\frac{3+4}{2},\frac{4+6}{2})=(5,5,0,5),\\
			\mathbf{d}_{13}=\frac{\mathbf{d}_1+\mathbf{d}_3}{2}=(\frac{1+4}{2},\frac{2+5}{2},\frac{3+2}{2},\frac{4+1}{2})=(6,0,6,6),\\
			\mathbf{d}_{23}=\frac{\mathbf{d}_2+\mathbf{d}_3}{2}=(\frac{2+4}{2},\frac{1+5}{2},\frac{4+2}{2},\frac{6+1}{2})=(3,3,3,0).
		\end{align*}
		We can see that $0$ and $5$ in $\mathbf{d}_{12}$ do not appear in $\mathbf{d}_{1}$ and  $\mathbf{d}_{2}$; $0$ and $6$ do not occur in 
		$\mathbf{d}_{1}$ and $\mathbf{d}_{3}$; $0$ and $3$ do not occur in  $\mathbf{d}_{2}$ and  $\mathbf{d}_{3}$. Therefore, the second condition of Definition \ref{def-NHSLR} holds. Then $\mathbf{D}$ is a $(7,3,4)$ NHSLR. \hfill $\square$	 
	\end{example}
	%
	%
	%
	%
	%
	%
	%
	%

	By leveraging the structure of cyclic Latin squares, a PDA with linear subpacketization can be constructed via an NHSLR. The main idea is as follows: Given a $(v,g,b)$ NHSLR, we first construct a Latin square $\mathbf{L}$ of order $v$ through cyclic shifting (to ensure that each integer in each row and column occurs at most once), and replicate it to obtain $g$ identical Latin squares. We then partition the entries of $\mathbf{L}$ into $v$ mutually disjoint orbits of size $v$. Every such orbit can be generated by a unique integer in $\mathbb{Z}_v$. In the $i$-th Latin square where $i\in[g]$, we retain the orbits generated by the integers in the $i$-th row of the NHSLR, replace all integers in the remaining orbits with the symbol ``*'', and vertically concatenate the resulting $g$ squares. This step guarantees that condition C$1$ holds and also that the entries in the orbits generated by the integers in the same column of NHSLR satisfy condition C$3$ in Definition~\ref{def-PDA}. Finally, to ensure that the entries in the orbits generated by the integers in the different columns of NHSLR satisfy condition C$3$ in Definition~\ref{def-PDA} too, we differentiate between orbit classes according to their associated columns.

\begin{figure}[h]
		\centering
		\includegraphics[scale=0.5]{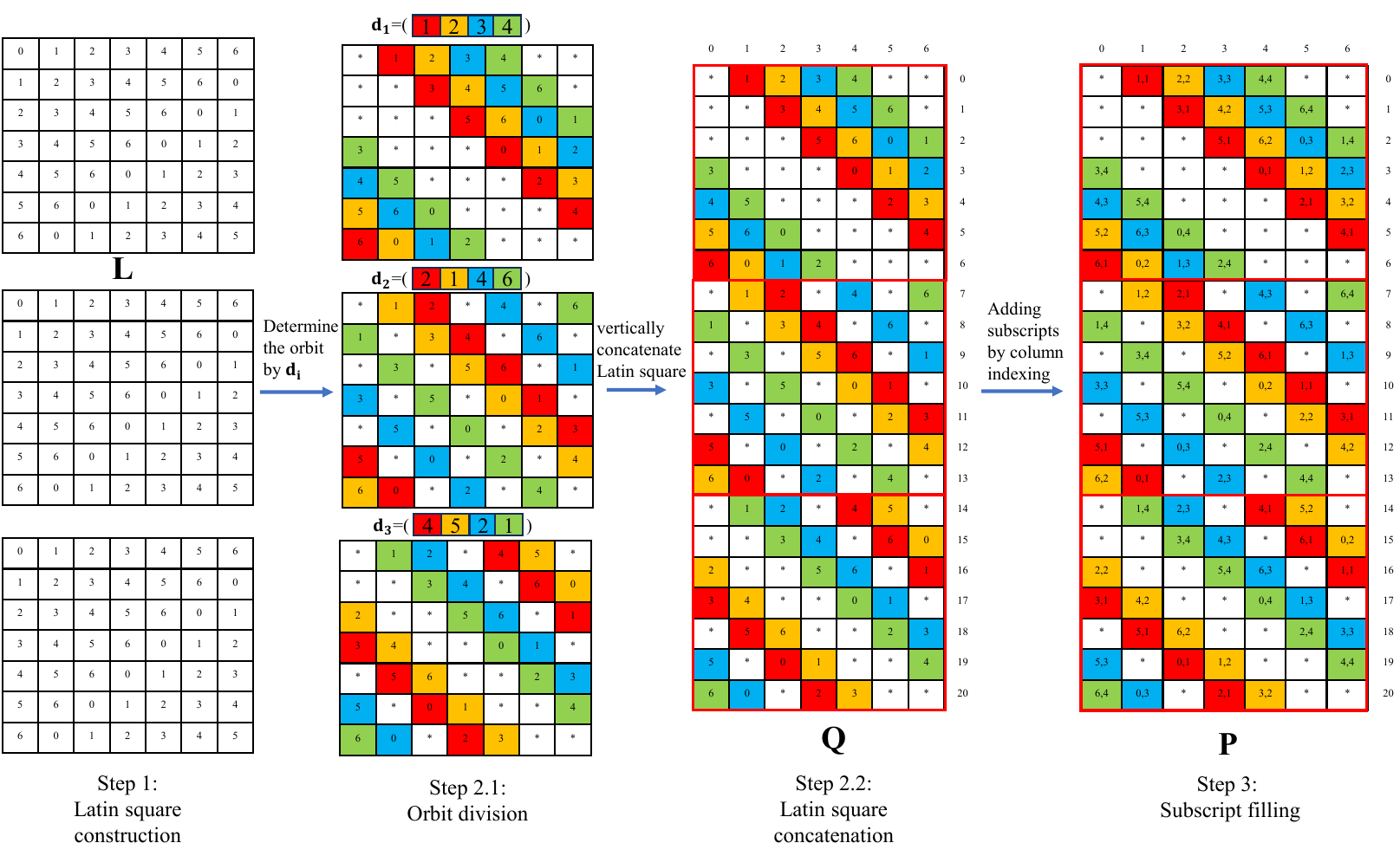}
		\caption{Flow diagram for constructing a PDA via $(7,3,4)$ NHSLR}
		\label{fig-example}
	\end{figure}

	As shown in Figure~\ref{fig-example}, we can obtain a $(7, 21, 9, 28)$ PDA via the $(v=7, g=3, b=4)$ NHSLR in Example~\ref{ex-NHSLR} by the following three steps.
	
	\begin{itemize}
		\item {\bf Step 1. Constructing Latin square.} We first construct a $7\times7$ classical Latin square using cyclic shifting: the first row is set to $0,1,\cdots,6$, and each subsequent row is generated by shifting the previous row one position to the right. As shown in Step $1$ of Figure~\ref{fig-example}, the Latin square can be defined as $\mathbf{L} = (l_{f,k})_{f,k\in \mathbb{Z}_7}$, where the entry $l_{f,k} = <f+k>_7$ for each $f,k\in \mathbb{Z}_7$. This structure ensures that each integer occurs at most once in every row and column, satisfying condition C$3$-a) of the PDA. Then we replicate $\mathbf{L}$ to obtain $g = 3$ identical Latin squares.
		\item {\bf Step 2. Orbit partition.} We then partition all entries of $\mathbf{L}$ into $7$ orbits. Specifically, for each integer $d \in \mathbb{Z}_7$, we define the orbit generated by $d$ as
		$$\text{Orbit-}d :=\{(f,k)\ | \ <k-f>_7 = d, f,k\in\mathbb{Z}_7\}.$$
		For instance, as listed in the first Latin square of Step $2.1$ in Figure~\ref{fig-example}, the Orbit-$1$ is $\{(0,1), (1,2), (2,3),(3,4),(4,5),(5,6),$ $(6,0)\}$ (i.e., the red grid). After partitioning the orbits, we process the generated Latin squares as follows:
		\begin{itemize}
			\item {\bf Step 2.1. Orbit retain and star replacement.} 
			We determine the orbits to be retained in each Latin square based on the integers in the $\mathbf{D}$. Let us consider the first Latin square. Since the first row of integers in $\mathbf{D}$ is $\mathbf{d}_1 = (1, 2, 3, 4)$, we retain Orbit-$1$, Orbit-$2$, Orbit-$3$, and Orbit-$4$ in the first Latin square, represented by red, yellow, blue, and green respectively. Then, we replace all integers in the remaining orbits (i.e.,  Orbit-$0$, Orbit-$5$, Orbit-$6$) with the symbol ``*''. Applying the same procedure to the other two Latin squares, we obtain three arrays composed of four orbits (red, yellow, blue, green) and the symbol ``*'' (see Step $2.1$ of Figure~\ref{fig-example}). We can check that every column of each array contains exactly $3$ star entries. 
			
			\item {\bf Step 2.2. Latin square concatenation.} We vertically concatenate the three arrays obtained in Step $2.1$ to form a $21\times 7$ array $\mathbf{Q}$, as shown in Step $2.2$ of Figure~\ref{fig-example}. We can see the array $\mathbf{Q}$ has exactly $Z = 9$ star entries in each column, satisfying condition C$1$ in Definition~\ref{def-PDA}. Furthermore, all entries within an orbit generated by integers from the same column of $\mathbf D$ (i.e., orbits of the same color) satisfy condition C$3$. For instance, we consider the red orbit generated by the integers in the first column of $\mathbf D$. We extract all rows and columns containing the entries $s=1$ to form a subarray $\mathbf{Q}^{(1)}_{\{1,2,4\}}$, as illustrated in Figure~\ref{fig-Half-sum}. It is not difficult to verify that the subarray $\mathbf{Q}^{(1)}_{\{1,2,4\}}$ satisfies the C$3$ condition. Similarly, we can verify that all entries within the same color orbit satisfy condition C$3$, and this is guaranteed by the non-half-sum property of the NHSLR. However, entries from orbits of different colors are not guaranteed to satisfy condition C$3$. This is because there is no strict non-half-sum constraint between integers in different columns of $\mathbf{D}$. Therefore, we need to add subscripts to the entries in the orbits to distinguish them.
			
			\begin{figure}[h]
				\centering
				\includegraphics[scale=0.5]{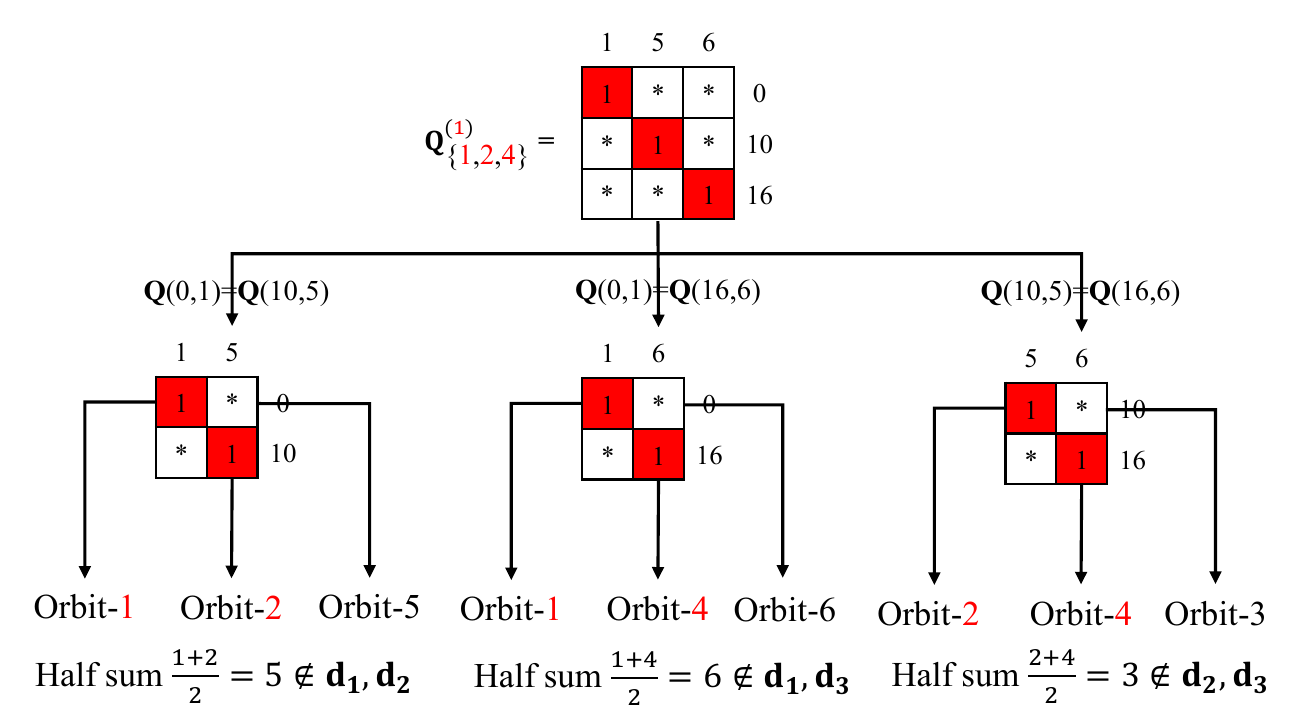}
				\caption{The subarray $\mathbf{Q}^{(1)}_{\{1,2,4\}}$ : Let us first consider two identical entries $\mathbf{Q}(0,1)$ and $\mathbf{Q}(10,5)$ in $\mathbf{Q}^{(1)}_{\{1,2,4\}}$ where $(0,1)\in \text{Orbit-}1,(10,5)\in \text{Orbit-}2$. The corresponding diagonal positions are $(0,5)$ and $(10,1)$, respectively. These two positions lie in the orbit generated by the half-sum $\frac{1+2}{2} = 5 \pmod{7}$ (i.e., Orbit-$5$). Since $5$ does not appear in either $\mathbf d_1$ or $\mathbf d_2$, the entries in this orbit are replaced by ``*'' in Step~$2.1$. Hence, the pair $\mathbf{Q}(0,1)$ and $\mathbf{Q}(10,5)$ satisfies condition C$3$. Similarly, we can verify that the pairs $\mathbf{Q}(0,1)$ and $\mathbf{Q}(16,6)$, as well as $\mathbf{Q}(10,5)$ and $\mathbf{Q}(16,6)$, also satisfy condition C$3$.  This is guaranteed by the non-half-sum property of the NHSLR.}
				\label{fig-Half-sum}
			\end{figure}

		\end{itemize}
		\item {\bf Step 3. Subscript assignment.}  	To ensure that all entries in $\mathbf{Q}$ satisfy the C$3$ condition defined in Definition~\ref{def-PDA}, we assign subscripts to the orbit entries to obtain $\mathbf{P}$. Specifically, for orbit generated by the $j$-th component of the each row vector, we add a uniform subscript $j$ to all entries in these orbits. As shown in the Step $3$ of Figure~\ref{fig-example}, we have added subscripts $1$, $2$, $3$, and $4$ to the entries in the red, yellow, blue, and green orbits, respectively. After adding the subscripts, we can check that any two identical non-star entries satisfy the condition C$3$. 
		
	\end{itemize}
	
	Through these three steps, we obtain a $(7,21,9,28)$ PDA, which can realize a coded caching scheme with memory ratio $M/N = 3/7$, subpacketization $F = gK = 21$, coded caching gain $g=3$, and load $R = 4/3$. Next, we consider the CWWC schemes with $K=7$. For $K=7$, a $(v=7,g=3,z=1)$ NHSDP can be obtained from \cite{CWWC}, which  can realizes a coded caching scheme with memory ratio $M/N=4/7$, coded caching gain $g=3$, and subpacketization $F=7$. Clearly, under the same coded caching gain, our scheme achieves a lower memory ratio than the CWWC scheme.

	In general, the mathematical representation of the above construction is as follows.
	\begin{construction}\rm\label{cons-PDA}
		Given a $(v,g,b)$ NHSLR $\mathbf{D}$, then a $vg\times v$ array $\mathbf{P}=(p_{f,k})_{f\in
		[0:vg-1],k\in [0:v-1]}$ is defined in the following way
		\begin{equation}\label{eq-cons-1}
			p_{f,k}=\begin{cases}
				(<f+k>_v,j),& \mbox{if\ } <k-f>_v = d_{\lfloor f/v \rfloor+1,j},\ \exists\ j\in[b];\\
				\ \ \ \ \ *, &\mbox{otherwise},
			\end{cases}
		\end{equation} where $d_{i,j}$ denotes the $(i,j)$-entry of $\mathbf{D}$.\hfill $\square$
	\end{construction}

	From the above, we can obtain the following relationship between NHSLR and PDA. The detailed proof is included in Appendix~\ref{proof-th-PDA}.
	\begin{theorem}[PDA via NHSLR]\label{th-PDA}\rm 
		Given a $(v, g, b)$ NHSLR, there is a $(v, vg, (v-b)g, bv)$ PDA which realizes a $(K = v,M,N)$ coded caching scheme with memory ratio $M/N=1- b/v$, subpacketization $F = vg$, and transmission load $R =b/g$.\hfill $\square$
	\end{theorem} 
	
	\begin{remark}\rm
		\label{re-even-case}  When $K$ is even, we can add a virtual user into the coded caching system, making the efficient number of users $K+1$ which can be solved by the  $(K+1,g,b)$ NHSLR in Theorem~\ref{th-PDA}. 
	\end{remark}
	By Theorem~\ref{th-PDA} and Remark \ref{re-even-case}, we can generate a PDA via NHSLR. Therefore, we will introduce the construction of NHSLR in the following.

	It is worth noting that a NHSDP can be used to generate a NHSLR. To demonstrate our construction method, let us take a $(v = 7, g = 3, z = 1)$ NHSDP $\mathfrak{D}=\{\mathcal{D}=\{1,2,4\}\}$ to construct a $(7,3,3)$ NHSLR as an example. First, take $\mathcal{D}$ as the vector $\mathbf{d}_1=(1,2,4)^\mathrm{T}$. Then by cyclically shifting $1$ and $2$ positions respectively, we can obtain two vectors $\mathbf{d}_2=(2,4,1)^\mathrm{T}$ and $\mathbf{d}_3=(4,1,2)^\mathrm{T}$. Finally, the following array can be obtained by concatenating $\mathbf{d}_1$, $\mathbf{d}_2$, and $\mathbf{d}_3$ horizontally.
	\begin{align*}
		\mathbf{D}=\begin{pmatrix}
			\mathbf{d}_1&
			\mathbf{d}_2&
			\mathbf{d}_3
		\end{pmatrix}=\begin{pmatrix}
			1&2&4\\
			2&4&1\\
			4&1&2
		\end{pmatrix}.	
	\end{align*} In fact, any NHSDP can be used to generate a NHSLR. That is the following result whose proof is included in Appendix~\ref{proof-th-NHSDP-construction-NHSLR}.
	\begin{theorem}[NHSLR via NHSDP]\rm\label{th-NHSDP-construction-NHSLR}
		For any $(v,g,z)$ NHSDP, there exists a $(v, g, b)$ NHSLR with $b = gz$.\hfill $\square$ 
	\end{theorem}
	
	It is not difficult to check that we cannot use the $(7,3,4)$ NHSLR from Example~\ref{ex-NHSLR} to generate a NHSDP with $v=7$ and $g=3$. This implies that the structure of NHSLR is weaker than that of NHSDP. The fundamental reason is as follows.

	\begin{remark}\rm\label{re-differences-NHSLR and NHSDP}
		According to Definitions \ref{def-NHSDP} and \ref{def-NHSLR}, a $(v,g,b)$ NHSLR is an array of size $g\times b$ over $\mathbb{Z}_v$, while a $(v,g,b)$ NHSDP contains $b$ blocks of $\mathbb{Z}_v$ each of which contains $g$ integers. Furthermore, integers may repeat within NHSLR, whereas each integer in NHSDP appears at most once. Regarding the non-half-sum property, NHSLR requires that no half-sum integer can appear in either of the two rows that generate it, whereas NHSDP requires that half-sum entries cannot appear in any block. Thus, the combinatorial structure of NHSLR is weaker than that of NHSDP. \hfill $\square$
	\end{remark}


	

	\section{Construction of Non-Half-Sum Latin Rectangles}\rm\label{sec-4}
	
	In this section, we propose a novel construction framework for NHSLR. Furthermore, we obtain an optimal solution by transforming the parameter selection into an integer optimization problem within this framework.
	
	We construct the NHSLR based on a diagonal matrix $\mathbf X$, where each integer in $\mathbf D$ is linearly represented by the diagonal entries of $\mathbf X$ together with a set of row and column coefficients. Specifically, we carefully design the integers in the diagonal matrix $\mathbf{X}$ such that integers with identical row or column coefficients are distinct, thereby satisfying the first Condition. When taking the midpoint vector of the coefficient vectors of any two rows, components cancel out in certain dimensions, confining the midpoint vector to a restricted region of the coefficient space. This ensures that the integer corresponding to the midpoint (i.e., the integer in the half-sum vector) does not coincide with any integer in the two original rows, thereby satisfying the second Condition. Finally, we select the smallest possible value of $v$ to preserve the original properties of $\mathbf{D}$ over $\mathbb{Z}_v$ while ensuring that the integers are fully utilized. Thus, we obtain an NHSLR. Here, we first introduce our general construction as follows.
	\begin{construction}\rm\label{cons-NHSLR}
		Let $m_1,m_2,\ldots,m_n$ be positive integers and define $\mathbf{X}=\mathrm{diag}(x_1,x_2,\ldots,x_n)$ where $x_1=1$ and $x_i=	\prod_{j=1}^{i-1}(m_j+1)$ for each $i\in[2:n]$. Let matrix $\mathbf{A}$ of size $g\times n$ have as rows all vectors in $\{-1,1\}^n$, where $g=2^n$, and let matrix $\mathbf{B}$ of size $n\times b$ have as columns all vectors in $[m_1] \times [m_2] \times \cdots \times [m_n]$, where $b=\prod_{i=1}^n m_i$. Construct the $g\times b$ array as
		$$\mathbf{D}=\mathbf{A}\mathbf{X}\mathbf{B}.$$

		
		\hfill$\square$
	\end{construction}

	Let us introduce the following example to further illustrate this idea.

	\begin{example}\rm
		\label{example-NHSLR}
		When $n=2$ and $m_1 = m_2 = 3$, we have $x_1=1$ and $x_2=m_1+1=4$. Then, we can obtain a diagonal matrix 
		\begin{align*}
			{\bf X}&=\begin{pmatrix}
				1 & 0\\
				0 &  4
			\end{pmatrix}.
		\end{align*}
		Next, using the vectors in  $\{-1,1\}^2$ and $[m_1]\times [m_2]=[3]^2$, we construct the following two matrix respectively.
		\begin{align*}
			{\bf A}&=\begin{pmatrix}
				-1 & -1\\
				-1 &  1\\
				1 & -1\\
				1 &  1
			\end{pmatrix},\qquad
			{\bf B}= \begin{pmatrix}
				1 & 1 & 1 & 2 & 2 & 2 & 3 & 3 & 3\\
				1 & 2 & 3 & 1 & 2 & 3 & 1 & 2 & 3
			\end{pmatrix}.
		\end{align*} Finally, we obtain the desired $2^2\times 3^2 = 4\times 9$ array 
		\begin{align*}
			\mathbf{D}=\mathbf{A}\mathbf{X}\mathbf{B}=
			\begin{pmatrix}
				-5 & -9 & -13 & -6 & -10 & -14 & -7 & -11 & -15\\
				3 &  7 & 11 & 2& 6 &  10 &  1 & 5& 9\\
				-3 & -7 & -11 & -2& -6 & -10 & -1 & -5& -9\\
				5 & 9 & 13 & 6 & 10 & 14 & 7 & 11 & 15
			\end{pmatrix}.
		\end{align*}We can check that all integers in the same row or column in $\mathbf{D}$ are distinct, satisfying the Latin property. Next, we verify the non-half-sum property. 
		For instance, consider the first and second rows of $\mathbf{D}$. Their half-sum vector is 
		
		\begin{align*}
		\left(\frac{3-5}{2},\frac{7-9}{2},\frac{11-13}{2},\frac{2-6}{2},\frac{6-10}{2},\frac{10-14}{2},\frac{1-7}{2},\frac{5-11}{2},\frac{9-15}{2}\right)=(1,-1,-1,-2,-2,-2,-3,-3,-3).
		\end{align*}
		It can be readily checked that none of the integers in this half-sum vector appears in either the first row or the second row of $\mathbf{D}$. 
		Similarly, any pair of distinct rows of $\mathbf{D}$ satisfies the non-half-sum property.

		We denote the union of the set of integers in the $i$-th row vector $\mathbf{d}_i$ of matrix $\mathbf{D}$ and the set of integers in the half-sum vector formed by $\mathbf{d}_i$ as $\mathcal{U}_i$. For instance, the set $\mathcal{U}_1=\{-15,-14,\ldots,0\}$ consists of $16$ consecutive integers, and the same holds for $\mathcal{U}_i$ for all $i\in[4]$. To ensure that the integers in $\mathcal{U}_i$ do not overlap on $\mathbb{Z}_v$, $v$ must be greater than or equal to $16$. Since $v$ is an odd number, we set $v = 17$. It then follows that
		\begin{align*}
			\mathbf{D}=
			\begin{pmatrix}
				12 & 8 & 4 & 11 & 7 & 3 & 10 & 6 & 2\\
				3 &  7 & 11 & 2& 6 &  10 &  1 & 5& 9\\
				14 & 10 & 6 & 15& 11 & 7 & 16 & 12& 8\\
				5 & 9 & 13 & 6 & 10 & 14 & 7 & 11 & 15
			\end{pmatrix},
		\end{align*}
		we can check that $\mathbf{D}$ over $\mathbb{Z}_v$ satisfies the definition of the NHSLR. Thus, we obtain a $(17, 4, 9)$ NHSLR.\hfill $\square$ 
	\end{example}
	

	Within this construction framework, for any positive integer $n$ and $n$ positive integers $m_1, m_2, \ldots, m_n$, we can obtain a $g \times b$ matrix $\mathbf{D}$, where $g = 2^n$ and $b = \prod_{i=1}^{n} {m_i}$. For any $i \in [g]$, denote by $\mathbf{d}_i$ the $i$-th row vector of $\mathbf{D}$. The union of the integer entries of $\mathbf{d}_i$ and those of its half-sum vector is given by
	\begin{align}
		\label{eq-union-a}
		\mathcal{U}_i=\{ a_{i,1} c_1x_1+a_{i,2}c_2x_2+\cdots+ a_{i,n}c_nx_n\mid c_j\in [m_j]\cup\{0\},j\in[n]\}\ .
	\end{align}
	Therefore, the number of integers in $\mathcal{U}_i$ is $\prod_{i=1}^{n} {(m_i+1)}$. For any odd integer $v \ge \prod_{i=1}^{n} (m_i + 1)$, the following theorem holds, and its proof is provided in Section~\ref{sec-6}.
	%
	%

	\begin{theorem}\rm
		\label{th-main-PDA}
		For any positive integer $n$ and $n$ positive integers $m_1, m_2, \ldots, m_n$, there exists a $(v, 2^n, \prod_{i=1}^{n} m_i)$ NHSLR with an odd integer $v \geq \prod_{i=1}^{n} {(m_i+1)} $, we can obtain a $2^n v$-division $(v, M, N)$ coded caching scheme with memory ratio $M/N = 1 - \frac{\prod_{i=1}^{n} m_i}{v}$, coded caching gain $g = 2^n$, and transmission load $R = \frac{\prod_{i=1}^{n} m_i}{2^n}$.\hfill $\square$ 
	\end{theorem}
	
	%
	
	By Theorem~\ref{th-main-PDA}, given the parameters $v$ and $n$ (recall that $K = v$ and $g = 2^n$), we aim to determine the values of $m_1, \ldots, m_n$ that minimize the memory ratio. In other words, for a fixed coded caching gain, our objective is to identify the coded caching scheme that requires the minimum memory ratio. To this end, we incorporate the constraint $v \ge \prod_{i=1}^{n} (m_i + 1)$ into the optimization problem of selecting $m_1, m_2, \ldots, m_n$. The problem can be formulated as follows:
	\begin{equation}\label{eq-optimization}
		\begin{aligned}
			\text{Problem 1.}  \ \ \  & \textbf{Maximize } \text{function } f=\prod_{i=1}^{n}m_i \\
			&\textbf{Constrains: } m_1,\ldots,m_n \in \mathbb{Z}^{+},\\
			&\quad\quad\quad \quad \quad \quad \prod_{i=1}^{n}(m_i+1) \leq v.
		\end{aligned}
	\end{equation}
	Problem~1 is an integer programming problem, which is known to be NP-hard. To simplify the problem, we first relax the constraint $m_1, m_2, \ldots, m_n \in \mathbb{Z}^{+}$ to $m_1, m_2, \ldots, m_n \in \mathbb{R}^{+}$, thereby transforming it into a  convex optimization problem. This relaxed problem is then solved using the method of Lagrange multipliers. Finally, we take the floor operation to the solution. For notational convenience, we define $q := K^{1/n}$. The following theorem presents a sub-optimal solution obtained via this approach, whose proof is given in Appendix~\ref{proof-th-main}. 
	\begin{theorem}\rm
		\label{th-main}
		A sub-optimal closed-form solution to Problem~1 is
		\begin{align}
			m_1=m_2=\cdots=m_n= \lfloor v^{1/n}-1\rfloor=\lfloor q-1\rfloor. 
			\label{eq-m1=m2}
		\end{align}
		Under the selection in~\eqref{eq-m1=m2}, the resulting PDA has the parameters 
		\begin{equation*}
			K=q^n,\ \ F=2^nq^n,\ \ Z=2^n(q^n-\lfloor q-1\rfloor^n),\ \ S=\lfloor q-1\rfloor^nq^n,
		\end{equation*}which realizes a coded caching scheme with memory ratio $M/N=1-\frac{\lfloor q-1\rfloor^n}{q^n}$ and the transmission load $R=(\frac{\lfloor q-1\rfloor}{2})^n$.
		
		\hfill $\square$ 
	\end{theorem}
	
	When $q$ is a positive odd integer satisfying $q \ge 3$, the solution in~\eqref{eq-m1=m2} becomes the optimal solution to Problem~1. By Theorem~\ref{th-main}, the following result is obtained.
	\begin{remark}[Optimal solution of Problem 1]\rm\label{re-optimal}
		When $q$ in Theorem~\ref{th-main} is an odd integer, we obtain a $(K = q^n, F = 2^n q^n, Z = 2^n (q^n - (q - 1)^n), S = (q - 1)^n q^n)$ PDA, which realizes a $(K, M, N)$ coded caching scheme, with memory ratio $M/N = 1 - (\frac{q-1}{q})^n$, 
		subpacketization $F = 2^n q^n$, and transmission load $R = \left(\frac{q - 1}{2}\right)^n$.\hfill $\square$ 
	\end{remark}
	Given a PDA, the authors in \cite{CJTY} showed that its corresponding conjugate PDAs can be obtained as follows.
	\begin{lemma}\cite{CJTY}\rm
		\label{permutations of PDA}
		Gvien a $(K,F,Z,S)$ PDA for some positive integers $K$, $F$, $Z$ and $S$ with $0< Z<F$, there exists a $(K,S,S-(F-Z),F)$ PDA. \hfill $\square$  
	\end{lemma}
	Based on Theorem~\ref{th-main} and Lemma~\ref{permutations of PDA}, the following PDA can be constructed.
	
	\begin{corollary}[Conjugate PDA of Theorem~\ref{th-main}]\rm
		\label{cor:conjugate}
		The conjugate PDA corresponding to the PDA given in Theorem~\ref{th-main} is a $(q^n, \lfloor q - 1 \rfloor^n q^n, \lfloor q - 1 \rfloor^n (q^n - 2^n), 2^n q^n)$ PDA, which realizes a coded caching scheme with memory ratio $M/N=1-(\frac{2}{q})^n$ and transmission load  $R=(\frac{2}{\lfloor q-1\rfloor})^{n}$.
	\end{corollary}
	\section{Performance Analysis}\label{sec-5}
	In this section, we will present theoretical and numerical comparisons with the existing schemes respectively to show the performance of our new scheme in Theorem~\ref{th-main}.
	\subsection{Theoretical Comparisons}
	Since the schemes in \cite{YTCC,CKSM,ASK,ZCW,AST} have specific parameters for the number of users and memory ratio, we do not compare with them in this subsection. Furthermore, the authors in \cite{WCLC} showed that their scheme includes the partition schemes from \cite{YCTC}, the hypergraph schemes from \cite{SZG}, and the OA schemes from \cite{CWZW} as special cases. Therefore, we only compare our scheme with those in \cite{MN,WCLC, WCWL, XXGL, CWWC}, respectively. 

		\subsubsection{Comparison with the MN scheme in \cite{MN}}
		When $t=q^n-\lfloor q-1\rfloor^n$, the MN scheme can be obtained with the memory ratio $\frac{M}{N}=1-\frac{\lfloor q-1\rfloor^n}{q^n}$, the subpacketization $F_{\text{MN}}={K\choose t}=\binom{q^n}{q^n-\lfloor q-1\rfloor^n}=\binom{q^n}{\lfloor q-1\rfloor^n}$, and the transmission load $R_{\text{MN}}=\frac{\lfloor q-1\rfloor^n}{q^n-\lfloor q-1\rfloor^n+1}$. 
		By Theorem~\ref{th-main}, our scheme is obtained with the same memory ratio, where $K=q^n$, $F=2^nq^n$, and the transmission load $R=(\frac{\lfloor q-1\rfloor}{2})^n$. Then the following results hold:
		\begin{align*}
			\frac{F_{\text{MN}}}{F}
			&=\frac{\binom{q^n}{\lfloor q-1\rfloor^n}}{2^nq^n}
			\approx\frac{q^{n\lfloor q-1\rfloor^n}}{2^nq^n} 
			>\frac{q^{n(q-2)^n}}{2^nq^n}
			=\frac{q^{n(q-2)^n-n}}{2^n}
			=\frac{K^{(q-2)^n-1}}{2^n},\\
			\frac{R_{\text{MN}}}{R}
			&=\frac{\frac{\lfloor q-1\rfloor^n}{q^n-\lfloor q-1\rfloor^n+1}}{(\frac{\lfloor q-1\rfloor}{2})^n}
			=\frac{2^n}{q^n-\lfloor q-1\rfloor^n+1}
			<\frac{2^n}{q^n-(q-1)^n+1}
			\approx \frac{1}{K}\cdot\frac{2^n}{1-(\frac{q-1}{q})^n}.
		\end{align*}
		Compared with the MN scheme, the multiplicative reduction in subpacketization of our scheme is proportional to $K^{(q-2)^n-1}/2^n$, which grows exponentially with the user number $K$ at a rate of $(q-2)^n$, while the increase in transmission is only $(q^n-\lfloor q-1\rfloor^n+1)/2^n$
		which decreases exponentially with base $2$ at a rate of $n$.
		
		\subsubsection{Comparison with the WCLC scheme in \cite{WCLC}}
		Let $m=z=n$, $k=q$, and $t=q-\lfloor q-1\rfloor$. The WCLC scheme in \cite{WCLC} has a user number $K=q^n$, with a memory ratio of $\frac{M}{N}=1-\frac{\lfloor q-1\rfloor^n}{q^n}$, a subpacketization of $F_{\text{WCLC}}=\Big\lfloor\frac{q-1}{\lfloor q-1\rfloor}\Big\rfloor^nq^{n-1}=q^{n-1}$, and a transmission load of $R_{\text{WCLC}}=(\frac{\lfloor q-1\rfloor}{\lfloor\frac{q-1}{\lfloor q-1\rfloor}\rfloor})^n=\lfloor q-1\rfloor^n$.
		From Theorem~\ref{th-main}, we can obtain a coded caching scheme with the same memory ratio and the same number of users, where the subpacketization is $F=2^nq^n$ and the transmission load is $R=(\frac{\lfloor q-1\rfloor}{2})^n$. Hence, we obtain
		\begin{align*}
			\frac{F_{\text{WCLC}}}{F}=\frac{q^{n-1}}{2^nq^n}=\frac{1}{2^nq},\quad 
			\frac{R_{\text{WCLC}}}{R}=\frac{\lfloor q-1 \rfloor^n}{(\frac{\lfloor q-1 \rfloor}{2})^n}=2^n.
		\end{align*}
		Compared to the WCLC scheme, our scheme increases the number of subpacketization by a factor of $2^nq$ while reducing the transmission load by a factor of $2^n$.
		\subsubsection{Comparison with the WCWL scheme in \cite{WCWL}}
	Let $K=q^n$ and $t=q^n-\lfloor q-1\rfloor^n$ in \cite{WCWL}. We will compare the WCWL scheme with the proposed scheme in the following cases:  
	\begin{itemize}
		\item If $(K-t+1)\mid K$ or $K-t=1$, we have 
		$F_{\text{WCWL}}=K=q^n,R_{\text{WCWL}}=\frac{\lfloor q-1\rfloor^n(\lfloor q-1\rfloor^n+1)}{2q^n}.$
		By Theorem~\ref{th-main}, we can obtain the following result
		\begin{align*}
			\frac{F_{\text{WCWL}}}{F}=\frac{q^n}{2^nq^n}=\frac{1}{2^n}, \ \ 
			\frac{R_{\text{WCWL}}}{R}=\frac{\frac{\lfloor q-1\rfloor^n(\lfloor q-1\rfloor^n+1)}{2q^n}}{(\frac{\lfloor q-1\rfloor}{2})^n}
			=\frac{2^n(\lfloor q-1\rfloor^n+1)}{2q^n}
			>\frac{2^n(q-2)^n}{2q^n}=\frac{1}{2}\left(2-\frac{4}{q}\right)^n.
		\end{align*}
		
		\item If $\langle K \rangle_{K-t+1}=K-t$, we can obtain the WCWL scheme with
		$F_{\text{WCWL}}
		=\left(2\left\lfloor\frac{q^n}{\lfloor q-1\rfloor^n+1}\right\rfloor+1\right)q^n,
		R_{\text{WCWL}}
		=\frac{\lfloor q-1\rfloor^n}{2\left\lfloor \frac{q^n}{\lfloor q-1\rfloor^n+1} \right\rfloor+1}.$
		By Theorem~\ref{th-main}, we have
		\begin{align*}
			\frac{F_{\text{WCWL}}}{F}&=\frac{\left(2\lfloor\frac{q^n}{\lfloor q-1\rfloor^n+1}\rfloor+1\right)q^n}{2^nq^n}
			=\frac{2\lfloor\frac{q^n}{\lfloor q-1\rfloor^n+1}\rfloor+1}{2^n},\\[0.5ex]
			\frac{R_{\text{WCWL}}}{R}&=\frac{\frac{\lfloor q-1\rfloor^n}{2\lfloor \frac{q^n}{\lfloor q-1\rfloor^n+1}\rfloor+1}}{(\frac{\lfloor q-1\rfloor}{2})^n}
			=\frac{2^n}{2\lfloor \frac{q^n}{\lfloor q-1\rfloor^n+1} \rfloor+1}
			>\frac{2^n(q-2)^n}{2q^n+(q-2)^n}
			>\frac{2^n(q-2)^n}{3q^n}
			=\frac{1}{3}\left(2-\frac{4}{q}\right)^n.
		\end{align*}
		
		\item Otherwise, the WCWL scheme can be expressed as
		$F_{\text{WCWL}}
		=2\lfloor\frac{q^n}{\lfloor q-1\rfloor^n+1}\rfloor q^n,
		R_{\text{WCWL}}
		=\frac{\lfloor q-1\rfloor^n}{2\lfloor \frac{q^n}{\lfloor q-1\rfloor^n+1} \rfloor}.$
		By Theorem~\ref{th-main}, it follows that
		\begin{align*}
			\frac{F_{\text{WCWL}}}{F}&=\frac{2\lfloor\frac{q^n}{\lfloor q-1\rfloor^n+1}\rfloor q^n}{2^nq^n}
			=\frac{2\lfloor\frac{q^n}{\lfloor q-1\rfloor^n+1}\rfloor}{2^n},\\[0.5ex]  
			\frac{R_{\text{WCWL}}}{R}&=\frac{\frac{\lfloor q-1\rfloor^n}{2\lfloor \frac{q^n}{\lfloor q-1\rfloor^n+1} \rfloor}}{(\frac{\lfloor q-1\rfloor}{2})^n}
			=\frac{2^n}{2\lfloor \frac{q^n}{\lfloor q-1\rfloor^n+1} \rfloor}
			>\frac{2^n(q-2)^n}{2q^n}
			=\frac{1}{2}\left(2-\frac{4}{q}\right)^n.
		\end{align*}
	\end{itemize}
	
	Since both our scheme and the WCWL scheme in \cite{WCWL} achieve linear subpacketization, the comparison primarily focuses on the transmission loads. The multiplicative reduction achieved by our scheme is at least $\frac{1}{3}\cdot(2-\frac{4}{q})^n$. Moreover, when $q>4$ and $n \geq \frac{\ln 3}{\ln (2-\frac{4}{q})}$, the inequality $\frac{1}{3}\cdot(2-\frac{4}{q})^n\geq 1$ always holds.
	
	\subsubsection{Comparison with the XXGL scheme in \cite{XXGL}}
	When the number of users $K=q^n$ in \cite{XXGL}, the XXGL scheme has the memory ratio $\frac{M}{N}=1-\frac{2}{q^n}\gg 1-\left(\frac{2}{q}\right)^n$, the subpacketization $F_{\text{XXGL}}=q^n$, and the transmission load $R_{\text{XXGL}}=\frac{q^n-1}{q^n}$. By Corollary~\ref{cor:conjugate}, our scheme has 
	$F=\lfloor q-1\rfloor^{n}q^n,R=\left(\frac{2}{\lfloor q-1\rfloor}\right)^{n}.$
	Thus,
	\begin{align*}
		\frac{F_{\text{XXGL}}}{F}=\frac{q^n}{\lfloor q-1\rfloor^{n}q^n}=\lfloor q-1\rfloor^{-n},\quad 
		\frac{R_{\text{XXGL}}}{R}=\frac{\frac{q^n-1}{q^n}}{\left(\frac{2}{\lfloor q-1\rfloor}\right)^{n}}
		\approx \frac{\lfloor q-1\rfloor^{n}}{2^n}.
	\end{align*}
	Compared with the XXGL scheme in \cite{XXGL}, although the memory ratio of our scheme is much smaller, the transmission load is reduced by approximately $\frac{\lfloor q-1\rfloor^{n}}{2^n}$ times, while the subpacketization increases only by a factor of $\lfloor q-1\rfloor^{n}$, which is much smaller than $K$. Therefore, our scheme exhibits a significant advantage in terms of transmission load compared with the XXGL scheme.

\subsubsection{Comparison with the CWWC scheme in \cite{CWWC}} 
	The author notes in \cite{CWWC} that when $q$ is an odd integer, the CWWC scheme achieves optimal performance with the users $K=q^n$, the memory ratio $M/N=1-(\frac{q-1}{q})^n$, the subpacketization $F_{\text{CWWC}}=q^n$, and the transmission load $R_{\text{CWWC}}=(\frac{q-1}{2})^n$. Since both our scheme and the CWWC scheme in \cite{CWWC} have linear subpacketization, we focus on the comparison of transmission loads. From Remark~\ref{re-optimal}, we can obtain a coded caching scheme with $K=q^n$, the memory ratio $M/N=1-(\frac{q-1}{q})^n$, the subpacketization $F=2^nq^n$, and the transmission load $R=(\frac{q-1}{2})^n$. This demonstrates that our scheme can achieve the optimal case of the CWWC scheme.
		
	In the general formulation of the CWWC scheme~\cite{CWWC}, the system parameters satisfy $K=v$, the transmission load is $R_{\text{CWWC}}=\prod_{i=1}^n m_i$, and the scheme requires $v \ge \prod_{i=1}^{n} (2m_i+1)$. In contrast, our scheme also has $K=v$ but only requires $v \ge \prod_{i=1}^{n} (m_i+1)$. Under the same $v$, our scheme supports more combinations of integer parameters $(m_1,\ldots,m_n)$, which implies that more coded caching schemes can be constructed for the same number of users $K=v$. In addition, since the lower bound of each factor $(2m_i + 1)$ is greater than the lower bound of $(m_i + 1)$, our scheme allows for a larger dimension $n$ under the same number of users $K=v$, leading to a larger multicast gain $g=2^n$. Under the PDA-based identity $R=\frac{(1-M/N)K}{g},$ a larger multicast gain directly yields a smaller transmission load. Thus, for the same $K$ and cache ratio $M/N$, our scheme can achieve a strictly lower transmission load than the CWWC scheme.

	\subsection{Numerical Comparisons}
	In this subsection, we present numerical comparisons between our scheme in Theorem~\ref{th-main} and existing schemes with different levels of subpacketization, including linear subpacketization~\cite{WCWL,ZCW,AST,ASK,XXGL,CWWC}, polynomial subpacketization~\cite{YTCC,CKSM}, and exponential subpacketization~\cite{MN,WCLC}. 
	Note that \cite{CWWC} shows that the ASK scheme in \cite{ASK} is a special case of the CWWC scheme, so it suffices to compare with the CWWC scheme. Moreover, since the XXGL scheme in \cite{XXGL} has a large memory ratio close to $1$, it is excluded from this subsection.
	
	\subsubsection{Comparison with the linear subpacketization scheme in \cite{WCWL,ZCW,AST,CWWC}}
	In Table~\ref{tab-numerical-1}, we provide numerical comparisons with the schemes in \cite{ZCW,AST,CWWC}. Compared with the ZCW scheme, our scheme in Theorem~\ref{th-main} supports more users, achieves a close memory ratio, and incurs a smaller transmission load. Compared with the AST scheme, our scheme attains a smaller memory ratio and a lower transmission load, while supporting a similar number of users at the cost of slightly higher subpacketization. Compared with the CWWC scheme, our scheme achieves a close memory ratio and the same number of users, while also reducing the transmission load. Although our scheme has a slightly higher subpacketization, we are more concerned about load since these are linear subpacketization schemes.
	\begin{table}[htbp!]
		\centering
		\caption{The numerical comparison between  the scheme in Theorem~\ref{th-main} and the schemes in \cite{ZCW,AST}}
		\label{tab-numerical-1}
		\begin{tabular}{|c|c|c|c|c|c|}
			\hline
			$K$   & $M/N$   & Scheme  & Parameters & Load   & Subpacketization \\ \hline 
			$32$ & $0.6875$ & ZCW scheme in  \cite{ZCW} & $(m,w)=(5,2)$ & $1.25$ & $32$  \\
			$33$ & $0.7576$ & Scheme in Theorem~\ref{th-main}& $(v,n)=(33,3)$ & $1$ & $264$  \\ \hline
			$128$ & $0.836$ & ZCW scheme in  \cite{ZCW} & $(m,w)=(7,2)$ & $2.625$ & $128$  \\ 
			$129$ & $0.876$ & Scheme in Theorem~\ref{th-main} & $(v,n)=(129,4)$ & $1$ & $2064$  \\ \hline
			$256$ & $0.8906$ & ZCW scheme in  \cite{ZCW} & $(m,w)=(8,2)$ & $2.41$ & $256$  \\ 
			$257$ & $0.8755$ & Scheme in Theorem~\ref{th-main} & $(v,n)=(257,5)$ & $1$ & $8224$  \\ \hline
			$512$ & $0.8359$ & ZCW scheme in  \cite{ZCW} & $(m,w)=(9,3)$ & $5.25$ & $512$  \\ 
			$513$ & $0.8421$ & Scheme in Theorem~\ref{th-main} & $(v,n)=(513,4)$ & $5.0625$ & $8208$  \\ \hline
			$512$ & $0.9297$ & ZCW scheme in  \cite{ZCW} & $(m,w)=(9,2)$ & $2.04$ & $512$  \\ 
			$513$ & $0.9376$ & Scheme in Theorem~\ref{th-main} & $(v,n)=(513,5)$ & $1$ & $16416$  \\ \hline
			$52$ & $0.28846$ & AST scheme in  \cite{AST} & $(r,k)=(2,13)$ & $9.25$ & $52$  \\ 
			$49$ & $0.26531$ & Scheme in Theorem~\ref{th-main}  & $(v,n)=(49,2)$ & $9$ & $196$  \\ \hline 
			$1332$ & $0.2515$ & AST scheme in  \cite{AST}  & $(r,k)=(2,333)$ & $249.25$ & $1332$  \\
			$1331$ & $0.2487$ & Scheme in Theorem~\ref{th-main}  & $(v,n)=(1331,3)$ & $125$ & $10648$  \\  \hline
			
			$2192$ & $0.2509$ & AST scheme in  \cite{AST}  & $(r,k)=(2,548)$ & $410.5$ & $2192$  \\
			$2199$ & $0.2142$ & Scheme in Theorem~\ref{th-main} & $(v,n)=(2199,3)$ & $216$ & $17592$  \\ \hline
			
			$2400$ & $0.50125$ & AST scheme in  \cite{AST} & $(r,k)=(3,300)$ & $149.625$ & $2400$  \\
			$2401$ & $0.460$ & Scheme in Theorem~\ref{th-main}  & $(v,n)=(2401,4)$ & $81$ & $38416$  \\  \hline
			
			$341$ & $0.8123$ & CWWC scheme in  \cite{CWWC} & $(v,n)=(341,3)$ & $8$ & $341$  \\
			$341$ & $0.7625$ & Scheme in Theorem~\ref{th-main}  & $(v,n)=(341,4)$ & $5.0625$ & $5456$  \\  \hline
			$713$ & $0.697$ & CWWC scheme in  \cite{CWWC} & $(v,n)=(713,3)$ & $27$ & $713$  \\
			$713$ & $0.641$ & Scheme in Theorem~\ref{th-main}  & $(v,n)=(713,4)$ & $16$ & $11408$  \\  \hline
			$1111$ & $0.7696$ & CWWC scheme in  \cite{CWWC} & $(v,n)=(1111,4)$ & $16$ & $1111$  \\
			$1111$ & $0.7813$ & Scheme in Theorem~\ref{th-main} & $(v,n)=(1111,5)$ & $7.5938$ & $35552$  \\ \hline
		\end{tabular}
	\end{table}
	
\begin{figure}[htbp!]
		\centering
		\begin{minipage}[b]{.5\textwidth}
			\centering
			\includegraphics[scale=0.6]{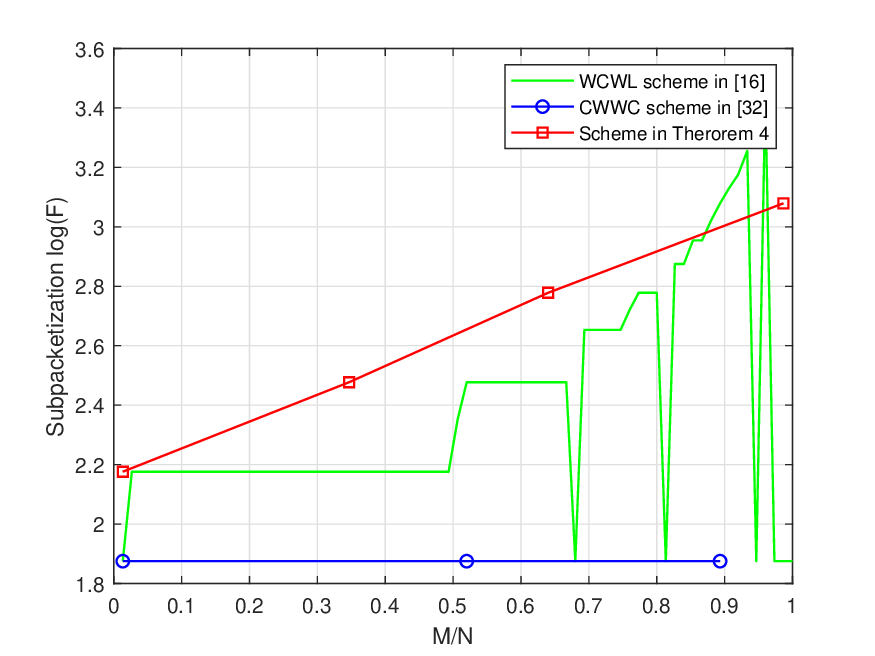}
			\caption{Memory ratio-subpacketization tradeoff for $K=75$} 
			\label{fig-com-75F}
		\end{minipage}
		\begin{minipage}[b]{.49\textwidth}
			\centering
			\includegraphics[scale=0.6]{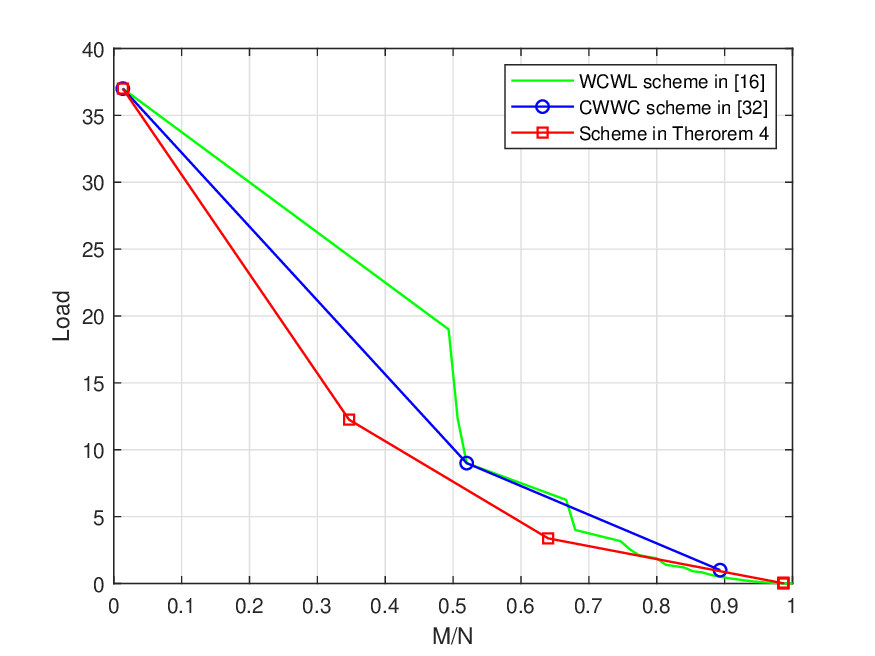}
			\caption{Memory ratio-load tradeoff for $K=75$} 
			\label{fig-com-75R}
		\end{minipage}
	\end{figure}

	Now, let us compare our scheme in Theorem~\ref{th-main} with the WCWL scheme in \cite{WCWL} and the CWWC scheme in \cite{CWWC} from the viewpoint of a line graph. When $K = 75$, our scheme (the red line), the CWWC scheme (the blue line), and the WCWL scheme (the green line) are shown in Figure~\ref{fig-com-75F} and Figure~\ref{fig-com-75R}. It can be seen that, under the same memory ratio, our scheme achieves a lower transmission load and a slightly higher subpacketization.
	
\begin{figure}[htbp!]
		\centering
		\begin{minipage}[b]{.5\textwidth}
			\centering
			\includegraphics[scale=0.6]{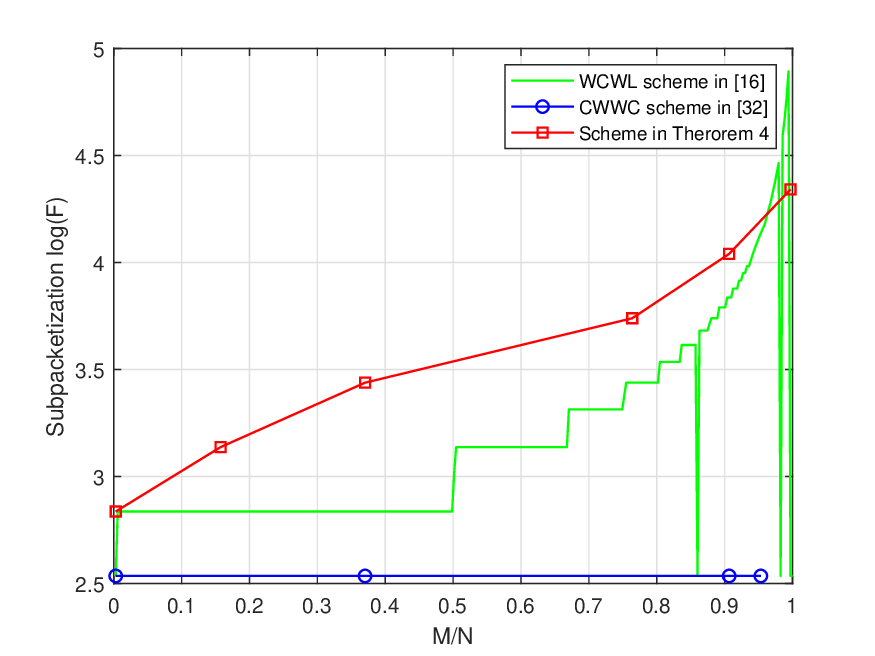}
			\caption{Memory ratio-subpacketization tradeoff for $K=343$} 
			\label{fig-com-343F}
			
		\end{minipage}
		\begin{minipage}[b]{.49\textwidth}
			\centering
			\includegraphics[scale=0.6]{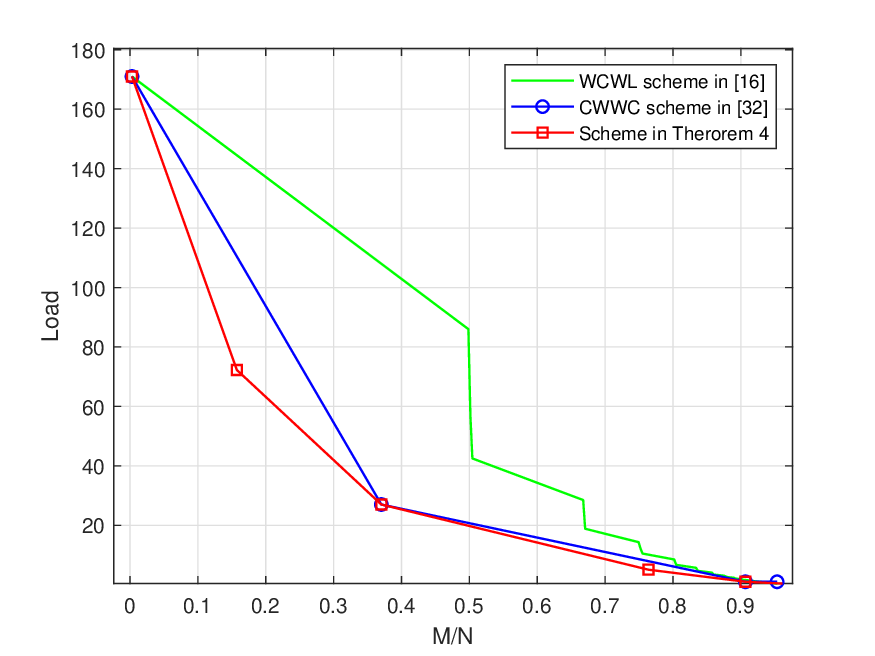}
			\caption{Memory ratio-load tradeoff for $K=343$} 
			\label{fig-com-343R}
		\end{minipage}
	\end{figure}

	Finally, we would like to point out that the CWWC scheme reaches its extreme when the number of users is a prime power, and the transmission load of our scheme is still smaller than or equal to that of the CWWC scheme, as shown in Figure~\ref{fig-com-343F} and Figure~\ref{fig-com-343R}.

	\subsubsection {Comparison with the  exponential subpacketization schemes in \cite{YTCC,CKSM}}
	Since the number of users in the schemes of \cite{YTCC,CKSM} is restricted to highly specialized integers derived from combinations, powers, or their products, it is difficult to directly illustrate the performance for a given value of $K$. Therefore, we compare our scheme with the YTCC and CKSM schemes by selecting specific user numbers and memory ratios, as listed in Table~\ref{tab-numerical-2}. Compared with the YTCC scheme, our scheme achieves similar memory ratios and supports more users, with a slightly larger transmission load but significantly smaller subpacketization. Compared with the CKSM scheme, our scheme provides similar transmission loads and much smaller subpacketization, while maintaining comparable user numbers and memory ratios.
	\begin{table}[htbp!]
		\centering
		\caption{The numerical comparison between the schemes in \cite{YTCC,CKSM} and the scheme in Theorem~\ref{th-main}}
		\label{tab-numerical-2}
		\begin{tabular}{|c|c|c|c|c|c|}
			\hline
			$K$   & $M/N$   & Scheme  & Parameters & Load   & Subpacketization \\ \hline

			$22$   & $0.364$  & YTCC scheme in \cite{YTCC}   & $(H,a,b,r)=(22,1,8,0)$  & $1.556$ & $319770$           \\
			$25$   & $0.36$   & Scheme in Theorem~\ref{th-main}  & $(v,n)=(25,2)$          & $4$ & $100$                   \\ \hline
			
			
			$136$  & $0.485$  & YTCC scheme in \cite{YTCC}   & $(H,a,b,r)=(17,2,7,1)$   & $7$    & $19448$            \\
			$127$ & $0.496$ & CKSM 2  scheme in \cite{CKSM}& $(q,k,m,t)=(2,7,6,1)$ & $9.143$ & $3.56E+09$   \\ 
			$125$  & $0.488$  & Scheme in Theorem~\ref{th-main}  & $(v,n)=(125,3)$             & $8$    & $1000$            \\ \hline
			
			$325$  & $0.354$  & YTCC scheme in \cite{YTCC}   & $(H,a,b,r)=(26,2,5,0)$ & $10$  & $65780$              \\
			$343$  & $0.370$  & Scheme in Theorem~\ref{th-main} & $(v,n)=(343,3)$               & $27$ & $2744$          \\\hline
			
			$231$  & $0.805$  & YTCC scheme in \cite{YTCC}   & $(H,a,b,r)=(22,2,12,0)$ & $0.4945055$  & $646646$     \\
			$255$ & $0.8784$ & CKSM 2 scheme in   \cite{CKSM} & $(q,k,m,t)=(2,8,4,1)$ & $2.0667$ & $97155$  \\ 
			$243$ & $0.8683$ & Scheme in Theorem~\ref{th-main} & $(v,n)=(243,5)$ & $1$ & $7776$  \\  
			\hline  
			$511$ & $0.8767$ & CKSM 2 scheme in    \cite{CKSM}& $(q,k,m,t)=(2,9,5,1)$ & $4.2$ & $788035$  \\ 
			$511$ & $0.8415$ & Scheme in Theorem~\ref{th-main} & $(v,n)=(511,4)$ & $5.0625$ & $8176$  \\ \hline
			
			$85$ & $0.247$ & CKSM 1  scheme in \cite{CKSM}& $(q,k,m,t)=(4,4,3,1)$ & $16$& $95200$   \\ 
			$85$ & $0.247$ & Scheme in Theorem~\ref{th-main} & $(v,n)=(85,2)$ & $16$  & $340$  \\ \hline
			
			$156$ & $0.1987$ & CKSM 1  scheme in \cite{CKSM}& $(q,k,m,t)=(5,4,3,1)$ & $31.25$ & $604500$   \\ 
			$151$ & $0.1987$ & Scheme in Theorem~\ref{th-main} & $(v,n)=(151,2)$ & $30.25$ & $604$   \\ \hline
			
			$156$ & $0.1987$ & CKSM 1 scheme in   \cite{CKSM} & $(q,k,m,t)=(5,4,3,1)$ & $31.25$ & $604500$  \\ 
			$169$ & $0.1479$ & Scheme in Theorem~\ref{th-main} & $(v,n)=(169,2)$ & $36$ & $676$  \\ \hline
			
			$255$ & $0.12157$ & CKSM 1  scheme in \cite{CKSM}& $(q,k,m,t)=(2,8,5,1)$ & $37.333$ & $8.10E+09$   \\ 
			$225$ & $0.1289$ & Scheme in Theorem~\ref{th-main} & $(v,n)=(225,2)$ & $49$ & $900$   \\ \hline
			
			$364$ & $0.332$ & CKSM 1  scheme in \cite{CKSM}& $(q,k,m,t)=(3,6,5,1)$  & $40.5$   & $4.51E+10$ \\ $343$ & $0.370$ & Scheme in Theorem~\ref{th-main} & $(v,n)=(343,3)$ & $27$  & $2744$   \\ \hline
		\end{tabular}
	\end{table}

	\subsubsection {Comparison with the  exponential subpacketization schemes in \cite{MN,WCLC}}
\begin{figure}[htbp!]
		\centering
		\begin{minipage}[t]{0.5\textwidth}
			\centering
			\includegraphics[scale=0.6]{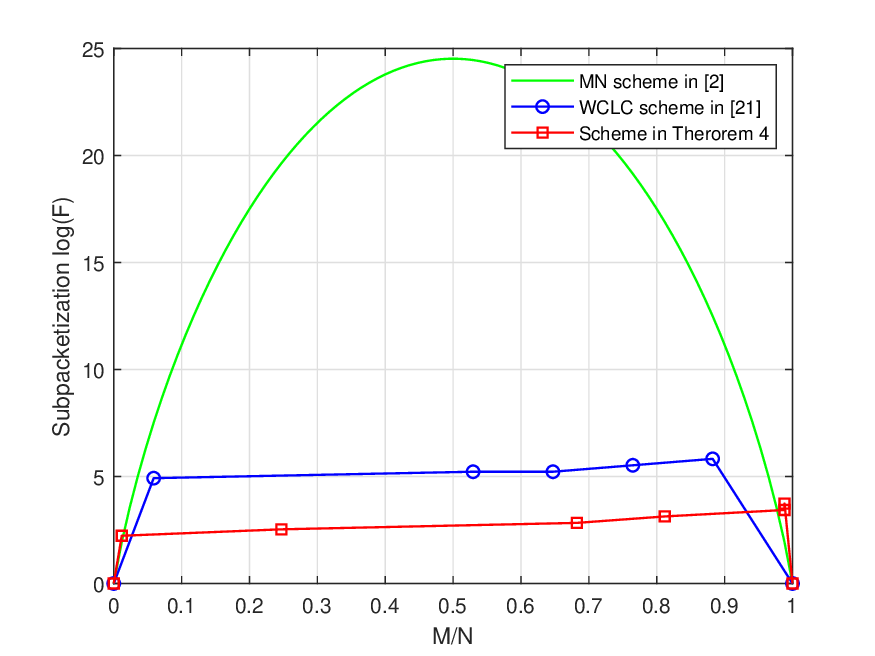}
			\caption{Memory ratio-subpacketization tradeoff for $K=85$} 
			\label{fig-comMN-F}
		\end{minipage}
		\begin{minipage}[t]{0.49\textwidth}
			\centering
			\includegraphics[scale=0.6]{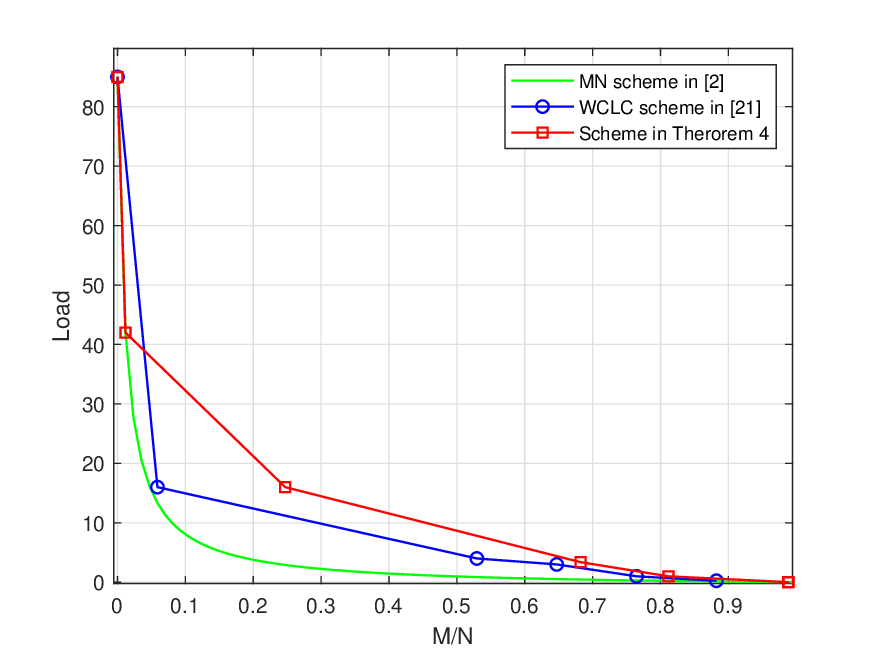}
			\caption{Memory ratio-load tradeoff for $K=85$} 
			\label{fig-comMN-R}
		\end{minipage}
	\end{figure}

	We first perform a direct comparison between the scheme in Theorem~\ref{th-main} and the MN and WCLC schemes, both of which have subpacketization levels that grow exponentially with the number of users, whereas our scheme achieves linear subpacketization. For $K = 85$, our scheme (red), the MN scheme (green), and the WCLC scheme (blue) are illustrated in Figures~\ref{fig-comMN-F} and \ref{fig-comMN-R}, according to Table~\ref{tab-Schemes} and Theorem~\ref{th-main}. Compared with the MN scheme (green line), our subpacketization (red line) is significantly smaller, while the transmission load is slightly larger. Compared with the WCLC scheme (blue line), our scheme (red line) achieves much smaller subpacketization while maintaining a transmission load that is close to or equal to that of the WCLC scheme.

	\section{PROOF OF THEOREM~\ref{th-main-PDA}}\label{sec-6}
			In this section, we prove that the matrix $\mathbf{D}$ over $\mathbb{Z}_v$ obtained by Construction~\ref{cons-NHSLR} satisfies the definition of an NHSLR when odd $v \ge \prod_{i=1}^{n} (m_i + 1)$. For any positive integer $i,i' \in [g]$ and $j,j' \in [b]$, by Construction~\ref{cons-NHSLR}, the integers of matrix $\bf D$ are given by
				$$d_{i,j} =a_{i,1}b_{j,1}x_1+a_{i,2}b_{j,2}x_2+\cdots+a_{i,n}b_{j,n}x_n \ \ \text{and} \ \ d_{i',j'} =a_{i',1}b_{j',1}x_1+a_{i',2}b_{j',2}x_2+\cdots+a_{i',n}b_{j',n}x_n,$$ where $a_{i,k} \in \{-1,1\}$ for all $i \in [g]$ and $k \in [n]$, and $b_{j,k} \in [m_k]$ for all $j \in [b]$ and $k \in [n]$.
				From Construction~\ref{cons-NHSLR}, for each $i\in[2:n]$, we have 
				$$x_{i}=\prod_{j=1}^{i-1}(m_j+1)=m_{i-1}x_{i-1}+x_{i-1}=\sum_{j=1}^{i-1}m_jx_j+x_1>\sum_{j=1}^{i-1}m_jx_j.$$
				Similarly, we can obtain $$v>\prod_{i=1}^{n}(m_i+1)>\sum_{i=1}^{n}m_ix_i.$$ 
				When $i = i'$ and $j \neq j'$, let $\ell = \max\{ e \mid b_{j,e} \neq b_{j',e},e \in [n]\}$. Then we have
				\begin{align}
					|d_{i,j}-d_{i',j'}|=&|\sum_{e=1}^{n}{a_{i,e}(b_{j,e}-b_{j',e})x_e} |=|\sum_{e=1}^{\ell}{a_{i,e}(b_{j,e}-b_{j',e})x_e} |
					=|a_{i,\ell}(b_{j,\ell}-b_{j',\ell})x_\ell+\sum_{e=1}^{\ell-1}{a_{i,e}(b_{j,e}-b_{j',e})x_e} |\nonumber\\
					\geq&|a_{i,\ell}(b_{j,\ell}-b_{j',\ell})x_\ell|-|\sum_{e=1}^{\ell-1}{a_{i,e}(b_{j,e}-b_{j',e})x_e}|
					\geq|a_{i,\ell}||(b_{j,\ell}-b_{j',\ell})|x_\ell-\sum_{e=1}^{\ell-1}{|a_{i,e}||(b_{j,e}-b_{j',e})|x_e}\nonumber\\
					\geq&|(b_{j,\ell}-b_{j',\ell})|x_\ell-\sum_{e=1}^{\ell-1}{m_ex_e}
					>|(b_{j,\ell}-b_{j',\ell})|x_\ell-x_\ell \geq 0. \label{eq-def-NHSLR-1-1}
				\end{align}
				On the other hand, we have the following upper bound
				\begin{align}
					|d_{i,j}-d_{i',j'}|=|\sum_{e=1}^{n}{a_{i,e}(b_{j,e}-b_{j',e})x_e}|
					\leq\sum_{e=1}^{n}{|a_{i,e}||(b_{j,e}-b_{j',e})|x_e}
					\leq\sum_{e=1}^{n}{m_ex_e}<v.
					\label{eq-def-NHSLR-1-2}
				\end{align}
				Combining inequalities~\eqref{eq-def-NHSLR-1-1} and~\eqref{eq-def-NHSLR-1-2}, we conclude that
				$$ 0 < |d_{i,j} - d_{i',j'}| < v,$$
				which implies that $d_{i,j} \not\equiv d_{i',j'} \pmod{v}$ when $i = i'$ and $j \neq j'$, i.e., the integers in the same row of matrix $\bf D$ are different modulo $v$.  
				
				When $j = j'$ and $i \neq i'$, let $\ell = \max\{e \mid a_{i,e} \neq a_{i',e},e \in [n]\}$. Then we have
				\begin{align}\label{eq-def-NHSLR-2-1}
					|d_{i,j}-d_{i',j'}|=&|\sum_{e=1}^{n}{(a_{i,e}-a_{i',e})b_{j,e}x_e} |=|\sum_{e=1}^{\ell}{(a_{i,e}-a_{i',e})b_{j,e}x_e} |
					=|(a_{i,\ell}-a_{i',\ell})b_{j,\ell} x_\ell+\sum_{e=1}^{\ell-1}{(a_{i,e}-a_{i',e})b_{j,e}x_e} |\nonumber\\
					\geq&|(a_{i,\ell}-a_{i',\ell})b_{j,\ell} x_\ell|-|\sum_{e=1}^{\ell-1}{(a_{i,e}-a_{i',e})b_{j,e}x_e}|
					\geq|(a_{i,\ell}-a_{i',\ell})|b_{j,\ell} x_\ell-\sum_{e=1}^{\ell-1}{|(a_{i,e}-a_{i',e})|b_{j,e}x_e}\nonumber\\
					\geq&2b_{j,\ell} x_\ell-\sum_{e=1}^{\ell-1}{2m_ex_e}
					>2b_{j,\ell} x_\ell -2x_\ell \geq 0.
				\end{align}
				Now let $\mathcal{S} = \{e \mid a_{i,e} \neq a_{i',e},e \in [n] \}$. Since for each $e \in [n]$, we have $a_{i,e}, a_{i',e} \in \{-1, 1\}$, it follows that
				\begin{align}
					|d_{i,j}-d_{i',j'}|=&|\sum_{e=1}^{n}{(a_{i,e}-a_{i',e})b_{j,e}x_e} |=|\sum_{e\in\mathcal{S}}{2a_{i,e}b_{j,e}x_e}|
					=2|\sum_{e\in\mathcal{S}}{a_{i,e}b_{j,e}x_e}|\label{eq-def-NHSLR-2-2_1}\\
					\leq&2\sum_{e\in\mathcal{S}}|a_{i,e}||b_{j,e}|x_e
					\leq2\sum_{e\in\mathcal{S}}m_ex_e
					\leq2\sum_{e=1}^{n}m_ex_e<2v.\label{eq-def-NHSLR-2-2_2}
				\end{align}
				From \eqref{eq-def-NHSLR-2-2_1}, we observe that $|d_{i,j} - d_{i',j'}|$ is an even integer, whereas $v$ is an odd number. Therefore, $|d_{i,j} - d_{i',j'}| \neq v$. 
				Combining the bounds in \eqref{eq-def-NHSLR-2-1} and \eqref{eq-def-NHSLR-2-2_2}, we obtain
				$
				0 < |d_{i,j} - d_{i',j'}| < 2v,
				$
				which implies that $d_{i,j} \not\equiv d_{i',j'} \pmod{v}$ when $j = j'$ and $i \neq i'$, i.e. the integers in the same column of matrix $\bf D$ are not equal modulo $v$.
				
				Now let us check the last condition of Definition~\ref{cons-NHSLR}. For any $i,i'\in [g]$ and $j\in [b]$, we define the half-sum of distinct integers $d_{i,j}$ and $d_{i',j}$ in the same column of matrix $\bf D$ as follows
				\begin{align} \label{eq-D1-D2-h}
					d_{ii',j}=\frac{d_{i,j}+d_{i',j}}{2}=\frac{a_{i,1}+a_{i',1}}{2}\cdot b_{j,1}x_1+\cdots+\frac{a_{i,n}+a_{i',n}}{2}\cdot b_{j,n}x_n.
				\end{align}
				We denote $\mathcal{S}=\{e \mid a_{i,e} = a_{i',e},e\in [n]\}$ and since $a_{i,e},a_{i',e}\in \{-1,1\}$ for each $e\in[n]$, then \eqref{eq-D1-D2-h} can be written as
				\begin{align} \label{eq-h-s}
					d_{ii',j}=\sum_{e\in\mathcal{S}}{a_{i,e}b_{j,e}x_e}=\sum_{e\in\mathcal{S}}{a_{i',e}b_{j,e}x_e}.
				\end{align}
				For any $j'\in [b]$, we have 
				\begin{align}\label{eq-a-s}
					d_{i,j'}=a_{i,1}b_{j',1}x_1+\cdots+a_{i,n}b_{j',n}x_n=\sum_{e\in\mathcal{S}}{a_{i,e}b_{j',e}x_e}+\sum_{e\in[n],e\notin \mathcal{S}}{a_{i,e}b_{j',e}x_e}.
				\end{align}
				From \eqref{eq-h-s} and \eqref{eq-a-s}, it follows that
				\begin{align}\label{eq-h-a}
					d_{i,j'}-d_{ii',j}=\sum_{e\in\mathcal{S}}{a_{i,e}(b_{j',e}-b_{j,e})x_e}+\sum_{e\in[n],e\notin \mathcal{S}}{a_{i,e}b_{j',e}x_e}.
				\end{align}
				Denote $\mathcal{T}=\{e \mid b_{j,e}\neq b_{j',e},e\in \mathcal{S}\}\cup\{e \mid e\in[n],e\notin\mathcal{S}\}$, we can write \eqref{eq-h-a} as
				\begin{align}
					d_{i,j'}-d_{ii',j}=\sum_{e\in \mathcal{T}}{a_{i,e}c_ex_e},\ \text{where}\ c_e=\begin{cases}
						b_{j',e}-b_{j,e},\ \ \ b_{j,e}\neq b_{j',e},e\in \mathcal{S};\\
						b_{j',e},\ \ \ \ \ \ \ \ \ \ \ e\in[n],e\notin\mathcal{S},
					\end{cases}\text{and}\ \  m_e\geq|c_e|\geq1.
				\end{align}
				Let $\ell=\max\{\mathcal{T}\}$, we have
				\begin{align}
					|d_{i,j'}-d_{ii',j}|=&|\sum_{e\in\mathcal{T}}{a_{i,e}c_ex_e} |=|a_{i,\ell} c_\ell x_\ell+\sum_{e\in\mathcal{T}\setminus\{\ell\}}{a_{i,e}c_ex_e} |
					\geq|a_{i,\ell} c_\ell x_\ell|-|\sum_{e\in\mathcal{T}\setminus\{\ell\}}{a_{i,e}c_ex_e}|\nonumber\\
					\geq&|a_{i,\ell}| |c_\ell| x_\ell-\sum_{e\in\mathcal{T}\setminus\{\ell\}}{|a_{i,e}||c_e|x_e}
					\geq|c_\ell| x_\ell-\sum_{e\in\mathcal{T}\setminus\{\ell\}}{m_ex_e}
					\geq|c_\ell| x_\ell-\sum_{e=1}^{\ell-1}{m_ex_e}\nonumber\\
					>&|c_\ell| x_\ell -x_\ell \geq 0.
					\label{eq-def-NHSLR-3-1}
				\end{align}
				On the other hand, we have the following upper bound
				\begin{align}
					|d_{i,j'}-d_{ii',j}|=|\sum_{e\in\mathcal{T}}{a_{i,e}c_ex_e} |
					\leq\sum_{e\in\mathcal{T}}{|a_{i,e}||c_e|x_e}
					\leq\sum_{e\in\mathcal{T}}{m_ex_e}
					\leq\sum_{e=1}^{n}{m_ex_e}<v.
					\label{eq-def-NHSLR-3-2}
				\end{align}
				Combining inequalities~\eqref{eq-def-NHSLR-3-1} and~\eqref{eq-def-NHSLR-3-2}, we conclude that
				$$ 0 < |d_{i,j'}-d_{ii',j}| < v,$$
				which implies that $d_{ii',j}\not\equiv d_{i,j'}$(mod $v$), i.e., $d_{ii',j}$ does not equal any integer in the $i$-th row of matrix $\bf D$ modulo $v$. Similarly, we can obtain $d_{ii',j}\not\equiv d_{i',j'}$ (mod $v$), which implies that $d_{ii',j}$ is also not equal to any integer in the $i'$-th row of matrix $\bf D$ modulo $v$. Then the proof is completed.
										
\section{Conclusion}\label{sec-7}
	Inspired by the NHSDP scheme, this paper proposes a novel combinatorial structure termed the Non-Half-Sum Latin  Rectangle (NHSLR). The PDA constructed using this combinatorial structure overcomes the disadvantage of the NHSDP scheme's overly restrictive linear subpacketization, while achieving a lower transmission load. According to theoretical and numerical comparisons, the proposed scheme outperforms existing schemes with linear subpacketization in terms of load reduction; in certain cases, it achieves lower load than some schemes with polynomial subpacketization; and in some parameter settings, it attains loads close to those of schemes with exponential subpacketization.
\appendices
\section{Proof of Theorem~\ref{th-NHSDP-construction-NHSLR}}
\label{proof-th-NHSDP-construction-NHSLR}
In general, for a $(v,g,z)$ NHSDP $\mathfrak{D} = \{\mathcal{D}_1, \dots, \mathcal{D}_z\}$, each $\mathcal{D}_k$ $(1 \leq k \leq z)$ can be represented as a column vector $\mathbf{d}_k$. For each $\mathbf{d}_k$, generate $g$ vectors by cyclically shifting its entries by $0,1,\dots,g-1$ positions. Concatenate all generated vectors horizontally to obtain a $g \times zg$ matrix $\mathbf{D}$. The resulting matrix $\mathbf{D}$ then forms a $(v,g,b)$ NHSLR with $b = gz$.
				
We now verify that $\mathbf{D}$ satisfies the properties of an NHSLR. For each block $\mathcal{D}_k \in \mathfrak{D}$, the $g$ cyclic shifts guarantee that every row of $\mathbf{D}$ contains all $g$ entries of $\mathcal{D}_k$ exactly once. Since each column of $\mathbf{D}$ is generated from a single block $\mathcal{D}_k$, the entries in each column are pairwise distinct. Moreover, because different blocks of the NHSDP are disjoint, all entries in each row of $\mathbf{D}$ are also distinct. Therefore, the first condition of the NHSLR is satisfied. Next, consider any two distinct rows of $\mathbf{D}$. Their entries at a given column position correspond to two distinct entries from the same block. By the non-half-sum property of the NHSDP, the half-sum of these two entries does not belong to any block of $\mathfrak{D}$, and therefore cannot appear in either of the two rows. Consequently, the half-sum vector of any two row vectors does not contain any entries that already appear in either of the two row vectors, thereby satisfying the second condition of an NHSLR. Hence, $\mathbf{D}$ is a $(v, g, zg)$ NHSLR.

\section{PROOF OF THEOREM~\ref{th-PDA} }
	\label{proof-th-PDA}
For any parameters $v$, $g$, and $b$, if a $(v, g, b)$ NHSLR exists, then a $(v, vg, (v-b)g, bv)$ PDA can be obtained via Construction~\ref{cons-PDA}. Assume that $\mathbf{D}$ is a $(v, g, b)$ NHSLR. For any $i \in [g]$, let $\mathbf{d}_i = (d_{i,1}, d_{i,2}, \ldots, d_{i,b})$ denote the $i$-th row vector of the matrix $\mathbf{D}$. According to Definition~\ref{def-PDA}, we first compute the number of star entries in each column.
For each column $k$ and each integer $x \in \mathbf{d}_{\lfloor f/v \rfloor + 1}$, there exists exactly one integer $f$ such that $\langle k - f \rangle_v = x \in \mathbf{d}_{\lfloor f/v \rfloor + 1}$.
This implies that there are exactly $bg$ integer entries in column $k$, i.e., $vg - bg$ star entries in column $k$. Therefore, we have $Z = vg - bg = (v - b)g$.
				
Next, for any two different entries $p_{f_1,k_1}$ and $p_{f_2,k_2}$ satisfying that $ p_{f_1,k_1}=p_{f_2,k_2}=(c,j)$, from \eqref{eq-cons-1} we have
				\begin{align} \label{eq-equal}
 	l_1=<k_1-f_1>_v=d_{\lfloor f_1/v \rfloor+1,j},\ l_2=<k_2-f_2>_v=d_{\lfloor f_2/v \rfloor+1,j}, \ \  \text{and}\ \ <k_1+f_1>_v=<k_2+f_2>_v. 
 \end{align} 
 If $f_1 = f_2$, then we have $k_1 = k_2$, which contradicts our hypothesis that $p_{f_1,k_1}$ and $p_{f_2,k_2}$ are distinct entries. If $k_1 = k_2$, then we must have either $f_1 = f_2$ or $f_1 = f_2 + nv$ where $n \in [-(g-1): g-1] \setminus \{0\}$. The case $f_1 = f_2$ again contradicts our hypothesis. If $f_1 = f_2 + nv$, then it implies $d_{\lfloor f_1 / v \rfloor + 1, j} = d_{\lfloor f_2 / v \rfloor + 1, j}$, which contradicts the first condition of Definition~\ref{def-NHSLR} that all entries in $\mathbf{D}$ are distinct in the same column. Therefore, each integer pair in $\mathbf{P}$ appears at most once in every row and every column. Next, consider the case where $f_1 \neq f_2$ and $k_1 \neq k_2$. First from \eqref{eq-equal}, we have $k_1=<f_1+l_1>_v, k_2=<f_2+l_2>_v$ and
 \begin{equation}\label{eq-k1k2}
 	<2f_1+l_1>_v=<2f_2+l_2>_v,\ \ \ \ \text{i.e.,}\ \ \ \ <l_1-l_2>_v=<2(f_2-f_1)>_v.
 \end{equation}
 Assume that the entry $p_{f_1,k_2}$ is not a star entry. Then  from \eqref{eq-equal} there exists a integer $j'\in[b]$ and $l_3\in \mathbb{Z}_v$ such that $l_3=<k_2-f_1>_v=d_{\lfloor f_1 / v \rfloor + 1,j'}$. This implies that $k_2=<l_3+f_1>_v$. Together with $k_2=<l_2+f_2>_v$ we have $<l_3-l_2>_v=<f_2-f_1>_v$. From \eqref{eq-k1k2} we have $<2(l_3-l_2)>_v=<l_1-l_2>_v$, i.e., $<2l_3>_v=<l_1+l_2>_v$. This implies that $$l_3=<\frac{l_1+l_2}{2}>_v=<\frac{d_{\lfloor f_1 / v \rfloor + 1,j}+d_{\lfloor f_2 / v \rfloor + 1,j}}{2}>_v\in\mathbf{d}_{\lfloor f_1 / v \rfloor + 1}$$ when $v$ is odd. This contradicts the NHSLR condition that any integer appearing in the half-sum of two row vectors in matrix $\mathbf{D}$ does not appear in either of the two original vectors. Similarly we can also show that $p_{f_2,k_1}=*$.
 
 Finally, we compute the number of distinct integer pairs in $\mathbf{P}$. For any integers $j\in [b]$, $c\in\mathbb{Z}_v$, when $v$ is odd, the following system of equations always have $g$ unique solutions
 \begin{equation*}
 	\begin{cases}
 		<k-f>_v=d_{\lfloor f / v \rfloor + 1,j},\\
 		<k+f>_v=c. 
 	\end{cases}
 \end{equation*}
 This means that each integer pair $(c, j)$ appears in $\mathbf{P}$ exactly $g$ times, and there are precisely $bv$ distinct integer pairs in total. Hence, the proof is complete.

 \section{Proof of Theorem~\ref{th-main}}
 \label{proof-th-main}
 Clearly, this is a convex optimization problem with $n$ real variables. The optimal solution can be obtained using the method of Lagrange multipliers. Specifically, we compute the partial derivatives of the function 
 \begin{align*}
 	\psi(m_1,\ldots,m_{n})= \prod_{i=1}^{n}m_{i}+\lambda\left(v-\prod_{i=1}^{n}(m_{i}+1)\right).
 \end{align*}
 With respect to the variable $m_1$, we have  
 \begin{align*}
 	\frac{\partial \psi(m_1,\ldots,m_{n})}{\partial m_1} = \prod_{i=2}^{n} m_i - \lambda \prod_{i=2}^{n}( m_i+1) = 0,
 \end{align*}which implies 
 \begin{align}
 	\label{eq-differ-1}
 	\prod_{i=2}^{n} m_i =\lambda \prod_{i=2}^{n}( m_i+1)
 \end{align}Similarly by partially differentiating of $\psi(m_1,\ldots,m_{n})$ with respect to the variable $m_2$, we have 
 \begin{align}
 	\label{eq-differ-2}
 	\frac{\partial \psi(m_1,\ldots,m_{n})}{\partial m_2} = m_1 \prod_{i=3}^{n} m_i - \lambda (m_1+1) \prod_{i=3}^{n}( m_i+1) = 0.
 \end{align}Substituting \eqref{eq-differ-1} into \eqref{eq-differ-2} and rearranging, we obtain $\frac{m_1}{m_2}=\frac{m_1+1}{m_2+1}$ which implies that $m_2=m_1$. Similarly, by partially differentiating $\psi(m_1,\ldots,m_{n})$ with respect to $m_3$, we have
 \begin{align}
 	\label{eq-differ-3}
 	\frac{\partial \psi(m_1,\ldots,m_{n})}{\partial m_3}=& m_1m_2 \prod_{i=4}^{n} m_i -\lambda (m_1+1)(m_2+1)\prod_{i=4}^{n}(m_i+1) = 0.
 \end{align}Substituting \eqref{eq-differ-1} into \eqref{eq-differ-3} and rearranging yields $\frac{m_1}{m_3}=\frac{m_1+1}{m_3+1}$ which implies $m_3=m_1$. 
 
 We now prove that $m_1=m_2=\cdots=m_{n}$. For any $k\in[n]$, taking the partial derivative of $\psi(m_1,\ldots,m_{n})$ with respect to $m_k$ and setting it to zero, we have
 
 \begin{align}
 	\label{eq-differ-k}
 	\frac{\partial \psi(m_1,\ldots,m_{n})}{\partial m_{k}}=&\prod_{i\in [n]\setminus\{k\}} m_i-\lambda \prod_{i\in [n]\setminus\{k\}}(m_{i}+1) = 0.
 \end{align} Substituting \eqref{eq-differ-1} into \eqref{eq-differ-k}, we obtain $\frac{m_1}{m_k}=\frac{m_1+1}{m_k+1}$ which implies $m_k=m_1$ where $k\in[n]$.
 
 From the above discussion, the optimal solution of Problem 1 is $m_1=m_2=\cdots=m_{n}$. From \eqref{eq-optimization} we have
 \begin{align*}
 	\prod_{i=1}^{n}( m_i+1)=(m_1+1)^n\leq v,
 \end{align*}which implies $m_1=m_2=\cdots=m_n\leq \sqrt[n]{v}-1$. So the length of each vector in NHSLR is $b\leq (\sqrt[n]{v}-1)^n$ and the memory ratio is at least $1-\frac{(\sqrt[n]{v}-1)^n}{v}$. 
\bibliographystyle{IEEEtran}
\bibliography{reference}
\end{document}